\documentclass{article}
\pdfoutput=1
\usepackage{arxiv}
\usepackage[square,numbers]{natbib}

\usepackage[utf8]{inputenc} 
\usepackage[T1]{fontenc}    
\usepackage{hyperref}       
\usepackage{url}            
\usepackage{booktabs}       
\usepackage{nicefrac}       
\usepackage{microtype}      
\usepackage{xcolor}         

\usepackage{amsfonts,amsmath,amssymb,amsthm}
\usepackage{graphicx}
\usepackage{subfigure}
\usepackage{enumitem}

\usepackage{appendix}

\DeclareMathOperator*{\argmax}{arg\!\max}

\newtheorem{proposition}{Proposition}

\newtheorem{definition}{Definition}
\newtheorem{lemma}{Lemma}
\newtheorem{theorem}{Theorem}
\newtheorem*{theorem*}{Theorem}

\newcommand{\mb}{\mathbb}
\newcommand{\mc}{\mathcal}

\newcommand{\R}{\mb{R}}
\newcommand{\bmat}[1]{\begin{bmatrix}#1\end{bmatrix}}
\newcommand{\X}{\mc{X}}
\newcommand{\A}{\mc{A}}
\newcommand{\supp}{\mathrm{supp}}

\newcommand{\onev}{\boldsymbol{1}}
\newcommand{\gda}{{\tt GDA}}
\newcommand{\ftrl}{{\tt FTRL}}

\newcommand{\email}[1]{\href{mailto:#1}{\texttt{#1}}}

\renewcommand{\headeright}{}
\renewcommand{\undertitle}{}

\title{Online Learning in Periodic Zero-Sum Games}

\author{%
  Tanner Fiez\thanks{Joint first authors}\\
  University of Washington\\
  Seattle, Washington\\
  \email{fiezt@uw.edu} \\
  \And
  Ryann Sim$^*$\\
  SUTD\\
  Singapore\\
  \email{ryann\_sim@mymail.sutd.edu.sg} \\
  \And
  Stratis Skoulakis$^*$\\
  SUTD\\
  Singapore\\
  \email{efstratios@sutd.edu.sg} \\
  \And
  Georgios Piliouras\thanks{Joint last authors}\\
  SUTD \\
  Singapore\\
  \email{georgios@sutd.edu.sg} \\
  \And
  Lillian Ratliff$^\dagger$\\
  University of Washington\\
  Seattle, Washington\\
  \email{ratliffl@uw.edu} \\

}

\date{}
\begin{document}

\maketitle
\begin{abstract}
  A seminal result in game theory is von Neumann's minmax theorem, which states that zero-sum games admit an essentially unique equilibrium solution. Classical learning results build on this theorem
  to show that online no-regret dynamics converge to an equilibrium in a time-average sense in zero-sum games. In the past several years, a key research direction has focused on characterizing the day-to-day behavior of such dynamics. General results in this direction show that broad classes of online learning dynamics are cyclic, and formally Poincar\'{e} recurrent, in zero-sum games. We analyze the robustness of these online learning behaviors in the case of periodic zero-sum games with a time-invariant equilibrium. This model generalizes the usual repeated game formulation while also being a realistic and natural model of a repeated competition between players that depends on exogenous environmental variations such as time-of-day effects, week-to-week trends, and seasonality. 
  Interestingly, time-average convergence may fail even in the simplest such settings, in spite of the equilibrium being fixed. In contrast, using novel analysis methods, we show that
  Poincar\'{e} recurrence provably generalizes despite the complex, non-autonomous nature of these dynamical systems.
\end{abstract}

\section{Introduction}
\label{sec:intro}
The study of learning dynamics in zero-sum games is arguably as old of a field  as game theory itself, dating back to the seminal work of Brown and Robinson~\cite{Brown1951,Robinson1951}, which followed shortly after the foundational minmax theorem of von Neumann~\cite{Neumann1928}. The dynamics of online no-regret learning algorithms~\cite{Cesa06, shalev2011online} are of particular interest in zero-sum games as they are designed with an adversarial environment in mind. 
Moreover, well known results imply that such dynamics converge in a time-average sense to a minmax equilibrium in zero-sum games~\cite{Cesa06,freund1999adaptive}.  

Despite the classical nature of the study of online no-regret learning dynamics in zero-sum games, the actual transient behavior of such dynamics was historically not as understood.
However, in the past several years this topic has gained attention with a number of works studying such dynamics
 in zero-sum  games (and variants thereof) with a particular focus on continuous-time analysis~\cite{piliouras2014optimization,piliouras2014persistent,mertikopoulos2018cycles, boone2019darwin,vlatakis2019poincare,perolat2020poincar,nagarajan2020chaos}. The unifying emergent picture is that the dynamics are ``approximately cyclic" in a formal sense known as Poincar\'{e} recurrence. 
 Moreover, these results have acted as fundamental building blocks for understanding the limiting behavior of their discrete-time variants~\cite{BaileyEC18,CP2019,2018arXiv180702629M,cheung2018multiplicative,bailey2020finite}.
 
 

Despite the plethora of emerging results regarding online learning dynamics in zero-sum games, an important and well motivated aspect of this problem has only begun to receive attention. 
\begin{quote}
\centering{\textit{How do online learning dynamics behave if the zero-sum game evolves over time?}}
\end{quote}
Clearly, the answer to this question depends critically on how the game is allowed to evolve.


 

\textbf{Problem Setting and Model.}
We study periodic zero-sum games with a time-invariant equilibrium, which is a class of games we formally define in Section~\ref{sec:prelims}. In a periodic zero-sum game, the payoffs that dictate the game are both $T$-periodic and zero-sum at all times. We consider both periodic zero-sum bilinear games on infinite, unconstrained strategy spaces and periodic zero-sum matrix games (along with network generalizations thereof) on finite strategy spaces.
The goal of this work is to evaluate the robustness of the archetypal online learning behaviors in zero-sum games, Poincar\'{e} recurrence and time-average equilibrium convergence, to this natural model of game evolution.

\textbf{Connections to Repeated Game Models.}
The time-evolving game model we study can be seen as a generalization of usual repeated game formulations. 
A time-invariant game is a trivial version of a  periodic game, in which case we recover the repeated static game setting. For a general periodic zero-sum game with period $T$, 
each stage game now is chosen according to a fixed length $T$ sequence of games, 
capturing interactions between the players with time-dependent payoffs. 

Periodic zero-sum games can also fit into the frameworks of multi-agent contextual games~\cite{sessa2020contextual} and dynamic games (see, e.g., \cite{bacsar1998dynamic}). In a multi-agent contextual game~\cite{sessa2020contextual}, the environment selects a context from a set before each round of play and this choice defines the game that is played. Periodic zero-sum games can be seen as a multi-agent contextual game where the environment draws contexts from the available set in a  $T$-periodic fashion with each context defining a zero-sum game with a common equilibrium. In the class of dynamic games, there is a game state on which the payoffs may depend that evolves with dynamics. Periodic zero-sum games can be interpreted as a dynamic game where the state transitions do not depend on the strategies of the players, the state is $T$-periodic, and the payoffs are completely defined by the state. We remark that the focus of existing work on contextual games and dynamic games is distinct from the questions investigated in this paper.

The periodic zero-sum game model allows us to capture competitive settings where exogenous environmental variations manifest in an effectively periodic/epochal fashion. This naturally occurs in market competitions where time-of-day effects, week-to-week trends, and seasonality can dictate the game between players. To illustrate this point, consider a competition between service providers that wish to maximize their users, while the total market size evolves seasonally over time. This 
evolution affects the utility functions, even if the fundamentals of the market,  and consequently the equilibrium, remain invariant. 

\textbf{Contributions and Approach.} In this paper, for the classes of periodic zero-sum bilinear games and periodic zero-sum polymatrix games with time-invariant equilibrium, we investigate the day-to-day and time-average behaviors of continuous-time gradient descent-ascent ({\gda}) and follow-the-regularized-leader ({\ftrl}) learning dynamics, respectively. This study highlights the careful attention that must be given to the dynamical systems in periodic zero-sum games which preclude standard proof techniques for Poincar\'{e} recurrence, while also revealing that intuition from existing results on static zero-sum games can be totally invalidated even by simple restricted examples in periodic zero-sum games.

\textbf{Contribution 1: Poincar\'{e} Recurrence.} A key technical challenge in this work is that the dynamical systems which emerge from learning dynamics in periodic zero-sum games correspond to \textit{non-autonomous} ordinary differential equations, whereas learning dynamics in static zero-sum games correspond to \textit{autonomous} ordinary differential equations. Consequently, the usual proof methods from static zero-sum games for showing Poincar\'{e} recurrence are insufficient on their own in periodic zero-sum games. We overcome this challenge by delicately piecing together properties of periodic systems to construct a discrete-time autonomous system that we are able to show is Poincar\'{e} recurrent. This approach allows to prove both  the {\gda} and {\ftrl} learning dynamics are Poincar\'{e} recurrent in the respective classes of periodic zero-sum games (Theorems \ref{thm:gdrec} \& \ref{thm:ftrlrec}). Finally, we show both periodicity and a time-invariant equilibrium are necessary for such results in evolving games (Proposition~\ref{prop:poincare_failure}).

\textbf{Contribution 2: Time-Average Strategy Equilibration Fails.}
Given that Poincar\'{e} recurrence provably generalizes from static zero-sum games to periodic zero-sum games, it may be expected that the time-average strategies in periodic zero-sum games converge to the time-invariant equilibrium as in static zero-sum games. Surprisingly, we show that counterexamples can be constructed to this intuition even in the simplest of periodic zero-sum games. In particular, we prove the negative result that the time-average {\gda} and {\ftrl} strategies do not necessarily converge to the time-invariant equilibrium in the respective classes of zero-sum games (Propositions \ref{prop:tagda} \& \ref{prop:tarep}). 


\textbf{Contribution 3: Time-Average Equilibrium Utility Convergence.}
Despite the negative result for time-average strategy convergence, in the special case of periodic zero-sum bimatrix games we are able to show a complimentary positive result on the time-average utility convergence. Specifically, 
 we show that the time-average utilities of the {\ftrl} learning dynamics converge to the 
 average of the equilibrium utility values of all the zero-sum games included in a single period of our time-evolving games. (Theorem \ref{thm:tavgconvergence}).

\textbf{Organization.}
 In Section~\ref{sec:prelims}, we formalize the classes of games that we study. We present 
 characteristics of dynamical systems 
 as they pertain to this work 
 in Section~\ref{sec:dynamical}. Section~\ref{sec:gda} and \ref{sec:ftrl} contain our results analyzing {\gda} and {\ftrl} learning dynamics in continuous and finite strategy periodic zero-sum bilinear and polymatrix games, respectively. We present numerical experiments in Section~\ref{sec:experiments} and finish with a discussion in Section~\ref{sec:discussion}. Proofs of our theoretical results are deferred to the appendix.

\section{Game-Theoretic Preliminaries}
\label{sec:prelims}

\subsection{Continuous Strategy Periodic Zero-Sum Games}
\label{sec:contstrat}
For continuous strategy periodic zero-sum games, we study periodic zero-sum bilinear games. We begin by formalizing zero-sum bilinear games and then define the periodic variant.

\textbf{Zero-Sum Bilinear Games.} Given a matrix $A\in \mb{R}^{n_1\times n_2}$, a zero-sum bilinear game on continuous strategy spaces can be defined by the max-min problem 
$\max_{x_1\in \mb{R}^{n_1}}\min_{x_2\in \mb{R}^{n_2}}x_1^{\top}Ax_2$.
Formally, the game is defined by the pair of payoff matrices $\{A, -A^{\top}\}$ and the action space of agents 1 and 2 are given by $\mb{R}^{n_1}$ and $\mb{R}^{n_2}$, respectively. 
Player 1 seeks to maximize the utility function $u_1(x_1,x_2)=x_1^{\top}Ax_2$ while player 2 optimizes the utility $u_2(x_1,x_2)=-x_2^{\top}A^{\top}x_1$. 
The game is zero-sum since for any $x_1\in \mb{R}^{n_1}$ and $x_2\in \mb{R}^{n_2}$, the sum of utility over each player is zero. 
For zero-sum bilinear games, a \emph{Nash equilibrium} corresponds to a joint strategy $(x_1^{\ast}, x_2^{\ast})$ such that for each player $i$ and $j\neq i$,
$u_i(x_i^\ast, x_{j}^{\ast})\geq u_{i}(x_i, x_{j}^\ast), \ \forall x_i\in \mb{R}^{n_i}.$
Note that $(x_1^{\ast}, x_2^{\ast})=(\mathbf{0}, \mathbf{0})$ is always a Nash equilibrium of a zero-sum bilinear game.

\textbf{Periodic Zero-Sum Bilinear Games.} We study the continuous-time {\gda} learning dynamics in a class of games we refer to as periodic zero-sum bilinear games. The key distinction from a typical static zero-sum bilinear game is that the payoff matrix is no longer fixed in this class of games. Instead, the payoff matrix may change at each time instant as long as game remains zero-sum and the continuous-time sequence of payoffs is periodic. The next definition formalizes this class of games.
\begin{definition}[Periodic Zero-Sum Bilinear Game]
A periodic zero-sum bilinear game is an infinite sequence of zero-sum bilinear games $\{A(t), -A(t)^{\top}\}_{t=0}^{\infty}$ in which the player set and strategy spaces are fixed and the payoff matrix is such that $A(t)=A(t+T)$ for a finite period $T$ and all $t\geq0$. Note that in such a game, $(0,0)$ is always a time-invariant Nash equilibrium. Furthermore, we assume that the dependence of the payoff entries on time is 
 smooth everywhere except for a finite set of points.
\label{def:bilinear}
\end{definition}

\subsection{Finite Strategy Periodic Zero-Sum Games}
\label{sec:finitestrat}
For finite strategy periodic zero-sum games, we analyze periodic zero-sum polymatrix games. In what follows we define a zero-sum polymatrix game, which is a network generalization of a bimatrix game, and then detail the periodic variant considered in this paper.

\textbf{Zero-Sum Polymatrix Games.}
An $N$-player polymatrix game is defined by an undirected graph $G = (V, E)$ where $V$ is the player set and $E$ is the edge set where a bimatrix game is played between the endpoints of each edge~\cite{cai2011minmax}. Each player $i \in V$ has a set of actions $\A_i = \{1,\ldots,n_i\}$ that can be selected at random from a distribution $x_i$ called a mixed strategy. The mixed strategy set of player $i \in V$ is the simplex in $\mb{R}^{n_i}$ denoted by $\X_i=\Delta^{n_i-1}=\{x_i\in \R^{n_i}_{\geq 0}:\ \sum_{\alpha \in \A_i} x_{i\alpha}=1\}$ where $x_{i\alpha}$ denotes the probability of action $\alpha \in \A_i$. 
The joint strategy space is denoted by by $\X = \Pi_{i \in V}\X_i$.

The bimatrix game on edge $(i,j)$ is described using a pair of matrices $A^{ij} \in \R^{n_i \times n_j}$ and $A^{ji} \in \R^{n_j \times n_i}$. 
The utility or payoff of agent $i \in V$ under the strategy profile $x \in \X$ is given by $u_i(x)=\sum\nolimits_{j:(i,j) \in E} x_i^\top A^{ij} x_j$ and corresponds to the sum of payoffs from the bimatrix games the player participates in. 
We further denote by $u_{i\alpha}(x) = \sum_{j:(i,j) \in E} (A^{ij} x_j)_{\alpha}$ the utility of player $i \in V$ under the strategy profile $x=(\alpha, x_{-i}) \in \X$ for $\alpha \in \A_i$.
The game is called zero-sum if $\sum_{i \in V}u_i(x) = 0$  for all $x \in \X$. Each bimatrix edge game is not necessarily zero-sum in a zero-sum polymatrix game.

A \emph{Nash equilibrium} in a polymatrix game is a mixed strategy profile $x^\ast\in \X$ such that for each player $i\in V$,
$u_i(x_i^\ast, x_{-i}^{\ast})\geq u_{i}(x_i, x_{-i}^\ast), \ \forall x_i\in \X_i$.
A Nash equilibrium is said to be an interior if $\supp(x_i^\ast)=\A_i\ \forall i\in V$ where $\supp(x^\ast_i)=\{\alpha \in \A_i:\ x_{i\alpha}>0\}$ is the support of $x^\ast_i\in \X_i$.

\textbf{Periodic Zero-Sum Polymatrix Games.}
We analyze the continuous-time {\ftrl} learning dynamics in a class of games we call periodic zero-sum polymatrix games. This class of games is such that the payoffs defined by the edge games evolve periodically. We consider that this periodic evolution is such that there is a common interior Nash equilibrium that arises in each zero-sum polymatrix game that arrives. 
The following definition formalizes the games we study on finite strategy spaces.
\begin{definition}[Periodic Zero-Sum Polymatrix Game]
A periodic zero-sum polymatrix game is an infinite sequence of zero-sum polymatrix games $\{G(t)=(V(t), E(t))\}_{t=0}^{\infty}$ in which the set of players, strategy spaces, and edges are fixed and each bimatrix game on an edge $(i, j)$ is such that  $A^{ij}(t)=A^{ij}(t+T)$ and $A^{ji}(t)=A^{ji}(t+T)$ for some finite period $T$ and all $t\geq0$. We assume there is a common interior Nash equilibrium $x^{\ast}\in \X$ of the polymatrix game $G(t)$ for all $t\geq 0$. Furthermore, we assume that the dependence of the payoff entries on time is 
  smooth everywhere except for a finite set of points.
\label{def:tvpolymatrix}
\end{definition}

\section{Preliminaries on Dynamical Systems }
\label{sec:dynamical}
We now cover concepts from dynamical systems theory that will help us analyze learning dynamics in periodic zero-sum games and prove Poincar\'e recurrence. 
Careful attention must be given to these preliminaries in this work since the dynamical systems we study are non-autonomous whereas typical recurrence analysis in the study of learning in games deals with autonomous dynamical systems.


\subsection{Background on Dynamical Systems}
We begin this section by providing dynamical systems background that is necessary both for defining Poincar\'e recurrence and sketching typical proof methods along with our approach.

\textbf{Flows.}  Consider an ordinary differential equation $\dot{x}=f(t,x)$ on a topological space $X$.
We can define the \emph{flow} $\phi:\R\times X\to X$ of a dynamical system $\dot{x}$, for which the following holds:
(i) $\phi(t,\cdot):X\to X$, often denoted $\phi^t:X\to X$, is a homeomorphism for each $t\in \R$,  (ii) $\phi(t+s,x)=\phi(t,\phi(s,x))$ for all $t,s\in\R$ and all $x\in X$,  (iii) for each $x\in X$, $\tfrac{d}{dt}|_{t=0}\phi(t,x)=f(t,x)$, and (iv) $\phi(t,x_0)=x(t)$ is the solution.

\textbf{Existence and Uniqueness.} We utilize Carath\'{e}odory's existence theorem to guarantee the existence of a flow for $\dot{x}$, even for some discontinuous functions $f$.
\begin{theorem*}[Carath\'{e}odory's existence theorem \cite{coddington1955theory, hale1980ordinary}]
Consider a differential equation $\dot{x}=f(t,x)$ on a rectangular domain $R = \{(t,y)\vert~|t-t_0| \leq a, |x-x_0|\leq b\}$. If $f$ satisfies the following conditions: 
\begin{enumerate}
    \item  $f(t,x)$ is continuous in $y$ for each fixed $t$,
    \item  $f(t,x)$ is measurable in $t$ for each fixed $y$,
    \item  there is a Lebesgue-integrable function $\displaystyle m:[t_{0}-a,t_{0}+a]\to [0,\infty )$ such that $\displaystyle |f(t,x)|~\leq~m(t)$ for all $\displaystyle (t,x)\in R$,
\end{enumerate}
then the differential equation has a solution. Moreover, if $f$ is also Lipschitz continuous, meaning $\vert f(t, x_1) - f(t,x_2) \vert \leq k(t)\vert x_1-x_2\vert$ with some Lebesgue-integrable function $k : [t_{0}-a,t_{0}+a]\to [0,\infty)$, then there exists a unique solution of the differential equation.
\end{theorem*}
In the settings we study, the above three conditions hold: Condition $1$ holds because for every fixed $t$, the dynamics we study (specifically \gda~and \ftrl) are continuous functions of their state space. Condition $2$ holds because the systems we study are finite and continuous almost-everywhere, and so by Lusin's theorem \cite{lusin1912proprietes} are measurable for each fixed $y$. Finally, Condition $3$ is always satisfied because the games we study always admit bounded orbits. Hence, it follows that a unique flow exists for all the dynamical systems studied in this paper.


\textbf{Conservation of Volume.} 
The flow $\phi$ of an ordinary differential equations is called \emph{volume preserving} if the volume of the image of any set $U \subseteq \R^d$ under $\phi^t$ is preserved, meaning that 
$\text{vol}(\phi^t(U)) = \text{vol}(U)$. 
\emph{Liouville's theorem} states that a flow is volume preserving if the divergence of $f$ at any point $x \in \R^d$ equals zero: that is, $\text{div}f(t,x)=\mathrm{tr}(Df(t,x))=\sum_{i=1}^d \frac{\partial  f(t,x)}{\partial x_{i}} = 0$.

We now transition to give general Poincar\'e recurrence statements along with discussion of how the results are usually applied in game theory before we outline our proof methods.

\subsection{Poincar\'{e} Recurrence in Autonomous Dynamical Systems}
A number of works in the past several years show that online no-regret learning dynamics are Poincar\'{e} recurrent in repeated static zero-sum games (see, e.g.,~\cite{piliouras2014optimization, mertikopoulos2018cycles, boone2019darwin}). The proof methods for deriving such results crucially rely on the static nature of the game for the reason that the learning dynamics amount to an autonomous dynamical system. Informally, the standard Poincar\'{e} recurrence theorem states that if an autonomous dynamical system preserves volume and every orbit remains bounded, almost all trajectories return arbitrarily close to their
initial position, and do so infinitely often~\cite{poincare1890probleme}; this property of a dynamical system is known as \emph{Poincar\'{e} recurrence}. Thus, proving the Poincar\'{e} recurrence of dynamics in repeated static zero-sum games is tantamount to verifying the volume preservation and bounded orbit properties. We now formalize the Poincar\'{e} recurrence theorem.

Given a flow $\phi^t$ on a topological space $X$, a point $x\in X$ is \emph{nonwandering} for $\phi^t$ if for each open neighborhood $U$ containing $x$, there exists $T>1$ such that $U\cap \phi^T(U)\neq \emptyset$. The set of all nonwandering points for $\phi^t$, called the \emph{nonwandering set}, is denoted $\Omega(\phi^t)$. 

\begin{theorem*}[Poincar\'{e} Recurrence for Continuous-Time Systems~\cite{poincare1890probleme}]\label{t:poincare}
If a flow preserves volume and has
only bounded orbits, then for each open set almost all orbits intersecting the set
intersect it infinitely often: if $\phi^t$ is a volume preserving flow on a bounded set $Z\subset \R^d$, then $\Omega(\phi^t)=Z$.
\end{theorem*}

In order to describe our proof methods in the following subsection for showing Poincar\'{e} recurrence in periodic zero-sum games, it will be useful for us to state an alternative formulation of the Poincar\'{e} recurrence theorem that is applicable to autonomous discrete-time systems.
\begin{theorem*}[Poincar\'{e} Recurrence for Discrete-Time Maps~\cite{barreira2006poincare}]\label{thm:poincare2}
Let $(X,\Sigma,\mu)$ be a finite measure space and let $\phi\colon X\to X$ be a measure-preserving map. 
For any $E\in \Sigma$, the set of those points $x$ of $E$ for which there exists $N\in\mathbb{N}$ such that $\phi^n(x)\notin E$ for all $n>N$ has zero measure. 
In other words, almost every point of $E$ returns to $E$. In fact, \emph{almost every point returns infinitely often}. That is, 
$\mu\left(\{x\in E: \exists \ N \text{ s.t. }
\phi^n(x)\notin E \text{ for all } n>N\}\right)=0. $ 
\label{thm:discreterecurrence}
\end{theorem*}

\subsection{Poincar\'{e} Recurrence in Periodic Dynamical Systems}
\label{sec:periodic}
The proof methods described in the last section for showing the Poincar\'{e} recurrence of dynamics in static zero-sum games cannot directly be applied to time-evolving zero-sum games as a result of the non-autonomous nature of the systems. In fact, we can construct time-evolving zero-sum games without both periodic payoffs and a time-invariant equilibrium, where
online learning dynamics are not Poincar\'{e} recurrent.\footnote{We formalize this statement in Proposition~\ref{prop:poincare_failure} of the following section.}
Despite this hurdle, we show that in the natural subclass of time-evolving games covered by periodic zero-sum games (periodic payoffs and a time-invariant equilibrium), we can develop proof methods to show the Poincar\'{e} recurrence of online learning dynamics. Given the previous claim regarding the possible non-recurrent nature of learning dynamics when there is not both periodic payoffs and a time-invariant equilibrium, this is perhaps the most general class of time-evolving zero-sum games with obtainable positive results in this direction.

We now give an overview of our approach, beginning by recalling properties of periodic systems.

\textbf{Periodic Systems and Poincar\'{e} Maps.}
A system $\dot{x}=f(t,x)$ is $T$-periodic if $f(t+T,x)=f(t,x)$ for all $(x,t)$. Let $\phi^t:\mb{R}^n\to\mb{R}^n$ denote the mapping taking $x\in \mb{R}^n$ to the value at time $t$. For a $T$-periodic system, $\phi^{T+s}=\phi^{s}\circ \phi^T$ so that $\phi^{kT}=(\phi^T)^k$ for any integer $k$. The mapping $\phi^T:\mb{R}^n\to\mb{R}^n$ is called the \emph{Poincar\'{e} map} or the \emph{mapping at a period}.  

If the differential equation is well-defined for all $x$ and has a solution
for all $t\in [0,T]$, then for each initial condition (where we have suppressed the dependence on $x_0$), the Poincar\'{e} map $\phi^T$ defines a discrete-time autonomous dynamical system $x^{+}=\phi^T(x)$. 
The learning dynamics we study in periodic zero-sum games form $T$-periodic dynamical systems. Thus, the discrete-time autonomous dynamical system $x^{+}=\phi^T(x)$ formed by the Poincar\'{e} map is key to
the analysis methods we pursue. In particular, our approach is to show that this system is Poincar\'{e} recurrent, which we then use to conclude that the original continuous-time non-autonomous system is Poincar\'{e} recurrent.

Given the previously presented Poincar\'{e} recurrence theorem for discrete-time maps, proving the Poincar\'{e} recurrence of the system $x^{+}=\phi^T(x)$ requires verifying the volume preservation and bounded orbit properties, which then implies the measure preserving property.
The following result states that if the divergence of a $T$-periodic vector field $f(x, t)$ is divergence free so that the flow $\phi^t$ is volume preserving, then the Poincar\'{e} map $\phi^T$ and the resulting discrete-time dynamical system $x^{+}=\phi^T(x)$ is also volume preserving.
\begin{theorem*}[Volume preservation for $T$-Periodic Systems~{\cite[3.16.B, Thm 2]{arnold2013mathematical}}]
If the $T$-periodic system $\dot{x}=f(t,x)$ is divergence-free, then $\phi^T$ preserves volume.
\label{thm:arnold}
\end{theorem*}
Similarly, if orbits of $\dot{x}=f(t, x)$ are bounded, then clearly the orbits of $x^{+}=\phi^T(x)$ are bounded. Hence, to show the system $x^{+}=\phi^T(x)$ is Poincar\'{e} recurrent, we prove $\dot{x}=f(t,x)$ has a divergence-free vector-field (equivalently, that the flow is volume preserving) and only bounded orbits. This will then be sufficient to conclude $\dot{x}=f(t,x)$ is Poincar\'{e} recurrent since the discrete-time system forms a subsequence of the continuous-time system.
\section{Gradient Descent-Ascent in Periodic Zero-Sum Bilinear Games}
\label{sec:gda}
This section focuses on the continuous-time {\gda} learning dynamics in periodic zero-sum bilinear games. The dynamics are such that each player seeks to maximize their utility by following the gradient with respect to their choice variable and are given  by 
\begin{align*}
\dot{x}_1&= A(t)x_2(t)\\
\dot{x}_2 &= -A^{\top}(t)x_1(t).
\end{align*}

\subsection{Poincar\'{e} Recurrence}
\label{sec:gda_recurrent}
The focus of this section is on characterizing the transient behavior of the continuous-time {\gda} learning dynamics in periodic zero-sum games. 
Specifically, we show the following result.
\begin{theorem}
The continuous-time {\gda} learning dynamics are Poincar\'{e} recurrent in any periodic zero-sum bilinear game as given in Definition~\ref{def:bilinear}.
\label{thm:gdrec}
\end{theorem}
Theorem~\ref{thm:gdrec} establishes that the recurrent nature of continuous-time {\gda} dynamics in static zero-sum bilinear games is robust to the dynamic evolution of the payoffs in periodic zero-sum bilinear games. 

Prior to outlining the proof steps for Theorem~\ref{thm:gdrec}, we elaborate on the claim from the previous section that without the periodicity property and a time-invariant equilibrium, such a result is unobtainable. In particular, we show the Poincar\'{e} recurrence of the {\gda} dynamics is not guaranteed without both properties by constructing counterexamples when only one of the properties holds.

\begin{proposition}
\label{prop:poincare_failure}
There exists time-evolving zero-sum games such that there is a time-invariant equilibrium or the payoffs are periodic (but not both simultaneously) in which the {\gda} dynamics are not Poincar\'{e} recurrent.
\end{proposition}

This proposition highlights the strength of our results regarding \gda, given that the assumptions needed to obtain them are more or less tight. 

We now outline the key intermediate results we prove to obtain Theorem~\ref{thm:gdrec}, following the techniques described in Section~\ref{sec:periodic}. 
For the {\gda} dynamics, we utilize the observation that the corresponding vector fields are divergence free to show that the learning dynamics are volume-preserving. We now state this result formally. 
\begin{lemma}
The {\gda} learning dynamics are volume preserving in any periodic zero-sum bilinear game as given in Definition~\ref{def:bilinear}.
\label{lemma:gdavolume}
\end{lemma}
We then proceed by showing that the {\gda} orbits are bounded by deriving a time-invariant function. This step relies on the fact that we have a time-invariant equilibrium.
\begin{lemma}
The function $\Phi(t)=\frac{1}{2}\big(x_1^\top(t) x_1(t)+x_2^\top(t) x_2(t)\big)$ is time-invariant. Hence, 
the {\gda} orbits are bounded in any periodic zero-sum bilinear game as given in Definition~\ref{def:bilinear}.
\label{lemma:gdabounded}
\end{lemma}
Given the volume preservation and bounded orbit characteristics of the continuous-time {\gda} learning dynamics in periodic zero-sum games, the proof of recurrence follows by applying the arguments described in Section~\ref{sec:periodic}.


\subsection{Time-Average Convergence}
\label{sec:gda_time_avg}
The Poincar\'{e} recurrence continuous-time {\gda} learning dynamics in periodic zero-sum bilinear games indicates that the system has regularities which couple the evolving players and evolving game despite the failure to converge to a fixed point. A natural follow-up question to the cyclic transient behavior of the dynamics is whether the long-run converges to a game-theoretically meaningful outcome.

We show that in periodic zero-sum bilinear games, the time-average of {\gda} learning dynamics may not converge to the time-invariant Nash equilibrium. To prove this, we consider a periodic zero-sum bilinear game with the action space of each player on $\mb{R}$ so that the evolving payoff simply rescales the vector field. We construct the payoff sequence so that the dynamics return back to the initial condition after a period of the game, while the time-average of the dynamics are not equal to the time-invariant equilibrium $(x_1^{\ast}, x_2^{\ast})=(\mathbf{0}, \mathbf{0})$.
Given the simplicity of this example, it effectively rules out hope to provide a meaningful time-average convergence guarantee in this class of games.
\begin{proposition}
There exists periodic zero-sum bilinear games satisfying Definition~\ref{def:bilinear} where the time-average strategies of the {\gda} dynamics fail to converge to the time-invariant equilibrium $(\mathbf{0}, \mathbf{0})$.
\label{prop:tagda}
\end{proposition}

\section{Follow-the-Regularized-Leader in Periodic Zero-Sum Polymatrix Games}
\label{sec:ftrl}
\label{subsec:replicator}
We now analyze continuous-time {\ftrl} learning dynamics in periodic zero-sum polymatrix games. Players that follow {\ftrl} learning dynamics in this class of games select a mixed strategy at each time that maximizes the difference between the cumulative payoff evaluated over the history of games and a regularization penalty. This adaptive strategy balances exploitation based on the past with exploration. 

Formally, the continuous-time {\ftrl} learning dynamics for any player $i\in V$ in a periodic zero-sum polymatrix game with an initial payoff vector $y_i(0)\in \mb{R}^{n_i}$ are given by
\begin{equation}
\begin{split}
y_i(t)&=\textstyle y_i(0)+\int_0^t \sum_{j:(i, j)\in E}A^{ij}(\tau)x_j(\tau) d\tau \\
x_i(t)&=\textstyle \argmax_{x_i\in \X_i}\{\langle x_i, y_i(t)\rangle-h_i(x_i)\}
\end{split}
\label{eq:fotrl}
\end{equation}
where $h_i:\X_i\rightarrow \mb{R}$ is a penalty term which encourages exploration away from the strategy which maximizes the cumulative payoffs in hindsight. We assume that the regularization function $h_i(\cdot)$ for each player $i\in V$ is continuous, strictly convex on $\X_i$, and smooth on the relative interior of every face of $\X_i$. These assumptions ensure the update $x_i(t)$ is well-defined since a unique solution exists.

Common {\ftrl} learning dynamics include the multiplicative weights update and the projected gradient dynamics. The multiplicative weights dynamics for a player $i\in V$ arise from the regularization function $h_i(x_i)=\sum_{\alpha\in \A_i}x_{i\alpha}\log x_{i\alpha}$ and correspond to the replicator dynamics. The projected gradient dynamics for a player $i\in V$ derive from the Euclidean regularization $h_i(x_i)=\frac{1}{2}\|x_i\|_2^2$.

To simplify notation, the {\ftrl} dynamics can equivalently be formulated as the following update
\begin{equation}
\begin{split}
y_i(t)&=\textstyle y(0)+\int_0^t v_i(x(\tau), \tau) d\tau \\
x_i(t)&=Q_i(y_i(t)).
\end{split}
\label{eq:fotrl2}
\end{equation}
Observe that we denote by $v_i(x, \tau)=(u_{i\alpha}(x, \tau))_{\alpha\in \A_i}$ the vector of each pure strategy $\alpha\in \A_i$ utility for agent $i\in V$ under the joint profile $x = (\alpha, x_{-i})\in \X$ at time $\tau\geq 0$. Moreover, $Q_i:\mathbb{R}^{n_i}\rightarrow \X_i$ is known as the choice map and defined as 
\[Q_i(y_i(t))=\textstyle \argmax_{x_i\in \X_i}\{\langle y_i(t),x_i\rangle-h_i(x_i)\}.\] 
In this notation, the utility of the player $i \in V$ under the joint strategy $x = (x_i, x_{-i})\in \X$ at time $t\geq 0$ is given by $u_i(x, \tau) = \langle v_i(x, \tau), x_i\rangle$. Observe that in our notation of utility we are now including the time index to make the dependence on the evolving game and payoffs explicit.

For any player $i\in V$ we denote by $h^\ast_i: \mb{R}^{n_i}\rightarrow \mb{R}$ the convex conjugate of the regularization function $h_i:\X_i\rightarrow \mb{R}$ which is given by the quantity $h^\ast_i(y_i(t))=\textstyle \max_{x_i\in \X_i}\{\langle x_i, y_i(t)\rangle-h_i(x_i)\}$.

\subsection{Poincar\'{e} Recurrence}
\label{sec:ftrl_recur}
We now focus on characterizing the transient behavior of the continuous-time {\ftrl} learning dynamics in periodic zero-sum polymatrix games. It is known that the continuous-time {\ftrl} learning dynamics are Poincar\'{e} recurrent in static zero-sum polymatrix games~\cite{mertikopoulos2018cycles}. The following result demonstrates that this characteristic holds even in games that are evolving in a periodic fashion with a time-invariant equilibrium, providing a broad generalization. 
\begin{theorem}
The {\ftrl} learning dynamics are Poincar\'{e} recurrent in any periodic zero-sum polymatrix game as given in Definition~\ref{def:tvpolymatrix}.
\label{thm:ftrlrec}
\end{theorem}
For the remainder of this subsection, we describe our proof methods. 
The general approach is that we prove the Poincar\'{e} recurrence of a transformed system using the techniques described in Section~\ref{sec:periodic}. This conclusion then allows us to infer the equivalent property for the original {\ftrl} system.


The utility differences for each player $i \in V$ and pure strategy $\alpha_i\in \A_i\setminus \beta_i$
evolve following the differential equation
\[\dot{z}_{i\alpha_i}=v_{i\alpha_i}(x(t),t)-v_{i\beta_i}(x(t),t).\]
Toward proving that this system is Poincar\'{e} recurrent, we show that the vector field $\dot{z}$ is divergence free and hence volume preserving.
\begin{lemma}
The dynamics defined by the system $\dot{z}$ are volume preserving in any periodic zero-sum polymatrix game as given in Definition~\ref{def:tvpolymatrix}.
\label{lem:ftrldiv}
\end{lemma}
We then construct a time-invariant function along the evolution of the system that is sufficient to guarantee that the orbits generated by the $\dot{z}$ dynamics are bounded. Recall that $x^{\ast}$ denotes the common, time-invariant interior Nash equilibrium of the periodic zero-sum polymatrix game.
\begin{lemma}
The function $\Phi(x^{\ast}, y(t))=\sum_{i\in V}\big(h_i^{\ast}(y_i(t)) - \langle x_i^{\ast},y_i(t) \rangle + h_i(x_i^{\ast})\big)$ is time-invariant. Hence, the orbits generated by the $\dot{z}$ dynamics are bounded in any periodic zero-sum polymatrix game as given in Definition~\ref{def:tvpolymatrix}.
\label{lem:boundorbit}
\end{lemma}
From this point, we follow the arguments from Section~\ref{sec:periodic} to conclude the $\dot{z}$ dynamics are Poincar\'{e} recurrent. 
Finally, we show that  Poincar\'{e} recurrence of the $\dot{z}$ system is sufficient to guarantee the  Poincar\'{e} recurrence of the {\ftrl} learning dynamics.

The proofs of the results in this section can be found in Appendix~\ref{app_sec:recurrence}.


\subsection{Time-Average Convergence}
\label{sec:ftrltavg}
A number of well-known properties of zero-sum bimatrix games fail to generalize to zero-sum polymatrix games. Indeed, fundamental characteristics of zero-sum bimatrix games include that each agent has a unique utility value in any Nash equilibrium and that equilibrium strategies are exchangeable. However,
\citet{cai2011minmax} show that neither of these properties are guaranteed in zero-sum polymatrix games.
Consequently, in general, time-average convergence to the set of equilibrium values in the utility and strategy spaces does not equate to the stronger notion of pointwise convergence.


For the reasons just outlined, we pursue a different notion of time-average convergence in periodic zero-sum polymatrix games. That is, we consider the subclass of periodic zero-sum bimatrix games (2-player periodic zero-sum polymatrix games) and show that the time-average utility of each agent converges to the time-average of the game values (that is, the unique utility the player obtains at any Nash equilibrium) over a period of the periodic game. 
\begin{theorem}
In periodic zero-sum bimatrix games satisfying Definition~\ref{def:tvpolymatrix}, if each player follows {\ftrl} dynamics, then the time-average utility of each player converges to the time-average over a period of the game equilibrium utility values. 
\label{thm:tavgconvergence}
\end{theorem}
Theorem~\ref{thm:tavgconvergence} paints a positive view of the time-average behavior of {\ftrl} learning dynamics in periodic zero-sum games. However, the following result demonstrates that much like in the case of {\gda} in periodic zero-sum bilinear games, the time-average strategies are not guaranteed to converge to the time-invariant Nash equilibrium.
\begin{proposition}
There exist periodic zero-sum bimatrix games satisfying Definition~\ref{def:bilinear} in which the time-average strategies of {\ftrl} dynamics fail to converge to the time-invariant Nash equilibrium.
\label{prop:tarep}
\end{proposition}
We prove the negative result in this proposition by constructing a simple counterexample that corresponds to a time-varying rescaling of Matching Pennies.

The proofs of the results in this section can be found in Appendix~\ref{app_sec:taverage}.

\section{Experiments}
\label{sec:experiments}
\begin{figure}[h]
 \centering
\begin{minipage}{.39\textwidth}
 \centering
  \includegraphics[width=.84\textwidth]{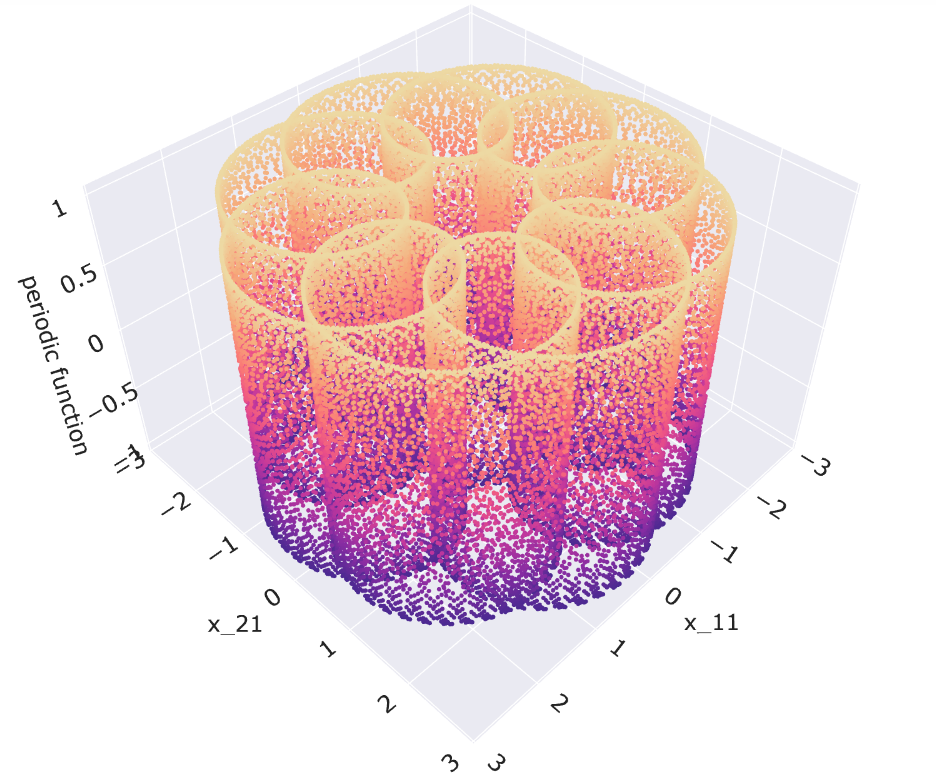}
  \caption{Bounded trajectories for a periodically rescaled Matching Pennies game updated using {\gda}.}\label{fig:gda_1}
\end{minipage}%
\hfill
\begin{minipage}{.58\textwidth}
  \centering
  {\includegraphics[width=\textwidth]{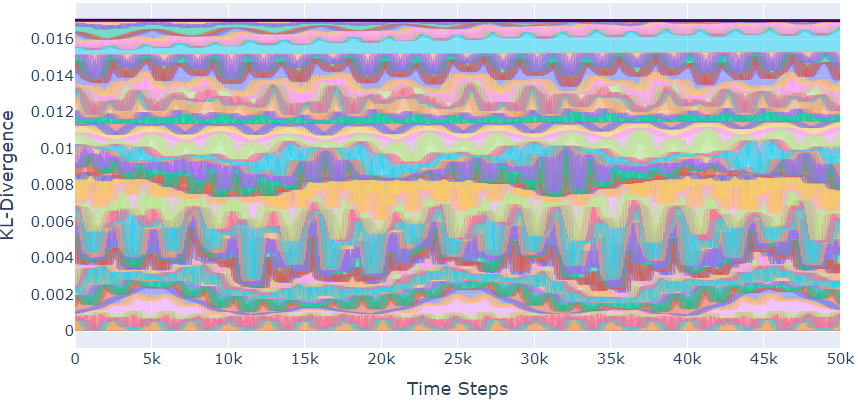}}
  \caption{Weighted sum of KL-divergences for a 64-player periodically rescaled Matching Pennies game. Note that despite the complicated trajectories of each player, the weighted sum of their divergences remains constant.}
  \label{fig:kldiv}
\end{minipage}
\end{figure}

In this section, we present several experimental simulations that illustrate our theoretical results. To begin, for continuous-time {\gda} dynamics we show that Poincar\'e recurrence holds in a periodic zero-sum bilinear game. We consider the ubiquitous Matching Pennies game with payoff matrix $ \bigl(\begin{smallmatrix}1 & -1\\ -1 & 1\end{smallmatrix} \bigr)$. We then use the following periodic rescaling with period $2\pi$:
 \[ \alpha(t) = \left\{
\begin{array}{ll}
      \sin(t) & 0\leq t \leq \frac{3\pi}{2} \\
      \left(\frac{2}{\pi}\right)(t\ \mathrm{mod}(2\pi)-2\pi) & \frac{3\pi}{2} \leq t\leq 2\pi
\end{array} 
\right. 
\]
When players follow the {\gda} learning dynamics, we see from Figure \ref{fig:gda_1} that the trajectories when plotted alongside the value of the periodic rescaling are bounded.  A similar experimental result holds in the case of {\ftrl} dynamics. In the supplementary material, we simulate replicator dynamics with the same periodic rescaling as in Figure \ref{fig:gda_1}. The trajectories in the dual/payoff space also remain bounded due to the invariance of the Kullback-Leibler divergences (KL-divergence). 

Lemmas \ref{lemma:gdabounded} and \ref{lem:boundorbit} describe functions $\Phi$ which remain time-invariant. In the case of replicator dynamics, $\Phi(t)$ is the sum of Kullback-Leibler divergences measured between the strategy of each player and their mixed Nash strategy $[1/2, 1/2]$. We simulated a 64-player polymatrix extension to the Matching Pennies game, where each agent plays against the opponent immediately adjacent to them, forming a 'toroid'-like chain of games. Furthermore, we randomly rescale each game with a different periodic function. Figure~\ref{fig:kldiv} depicts the claim presented in the lemmas: although each agent's specific divergence 
term $\mathrm{KL}(x^*_i \| x_i(t))$  fluctuates, the sum $\sum_{i \in V}\mathrm{KL}(x^*_i \| x_i(t))$ remains constant.

\begin{figure}[h]
\centering
\begin{tabular}{ccccccc}
  \includegraphics[width=0.125\textwidth]{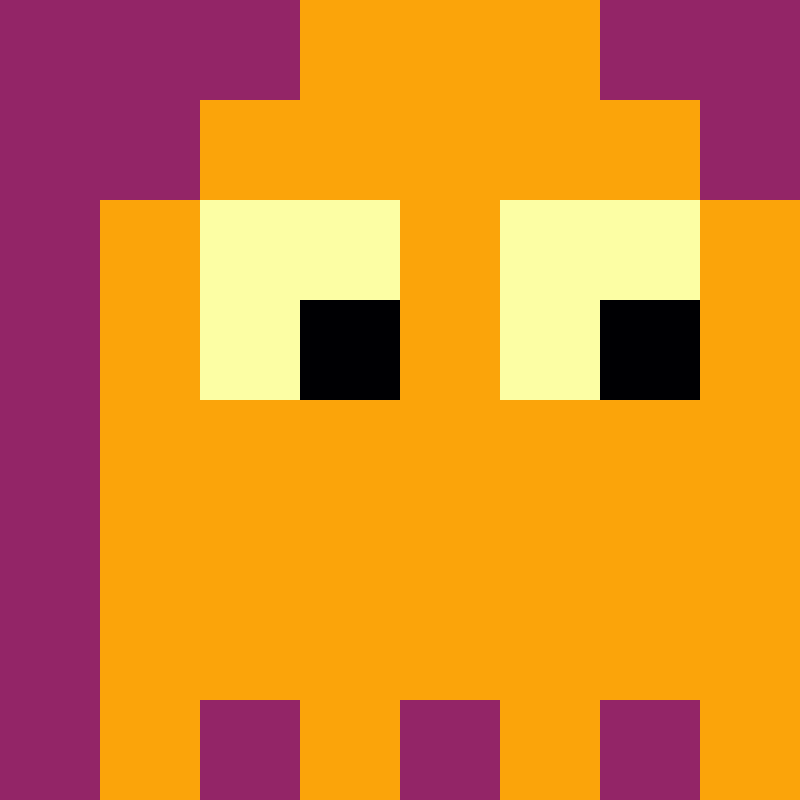} &  
 \includegraphics[width=0.125\textwidth]{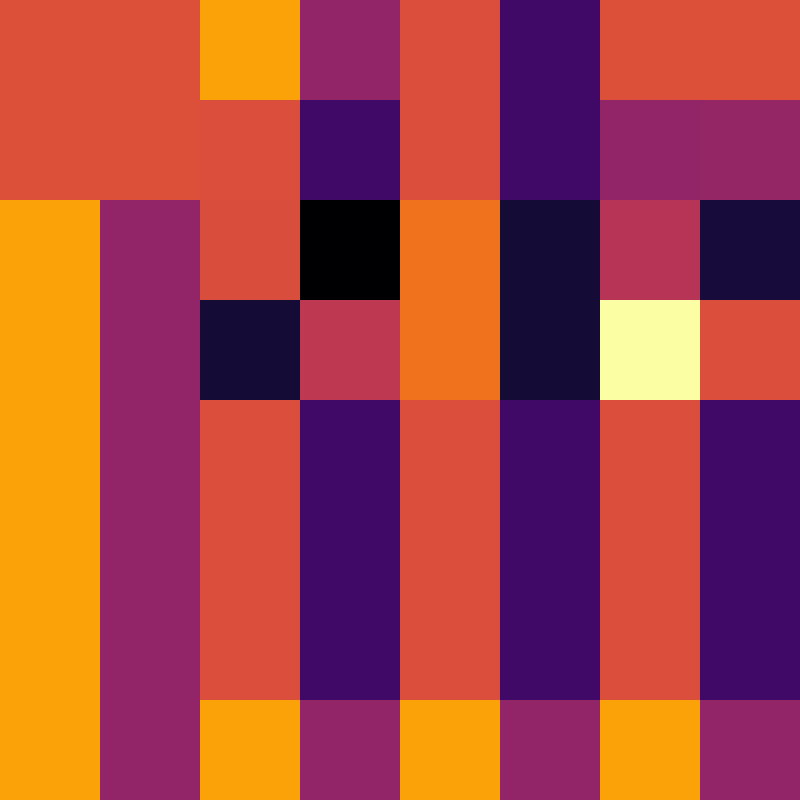} & \includegraphics[width=0.125\textwidth]{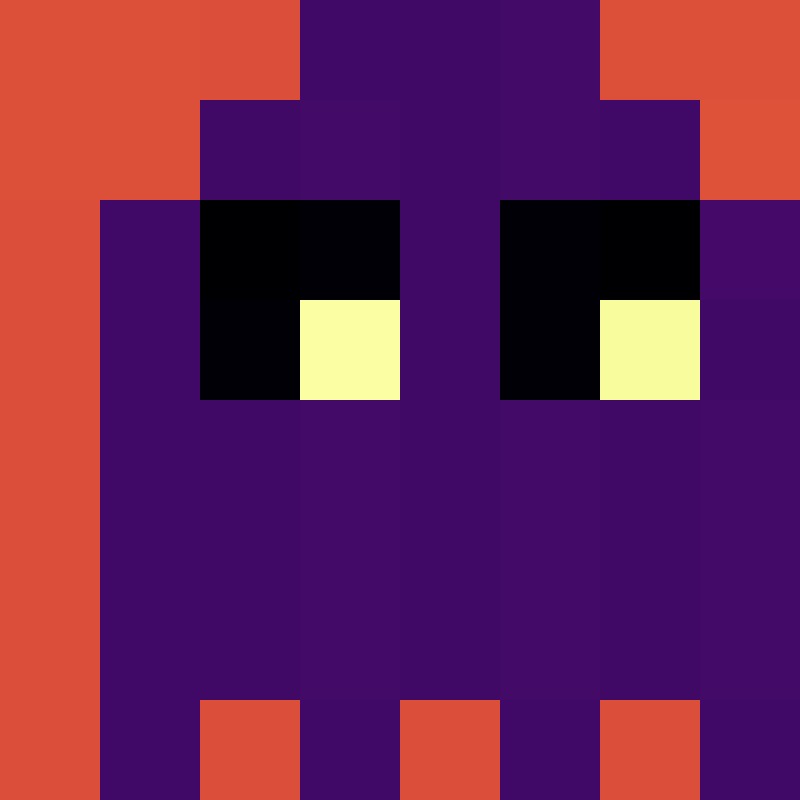} &   \includegraphics[width=0.125\textwidth]{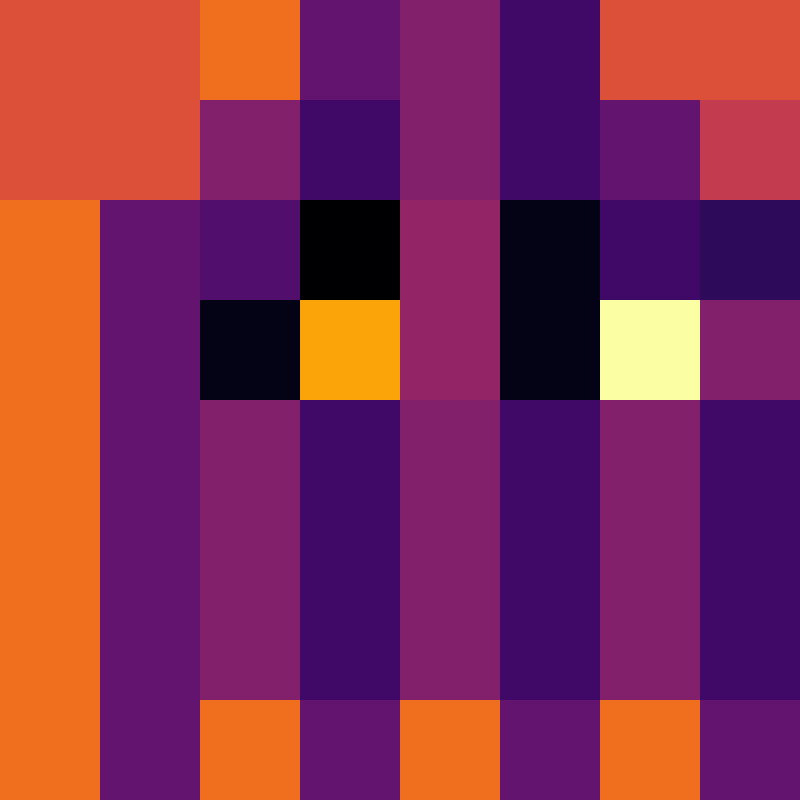} &
\includegraphics[width=0.125\textwidth]{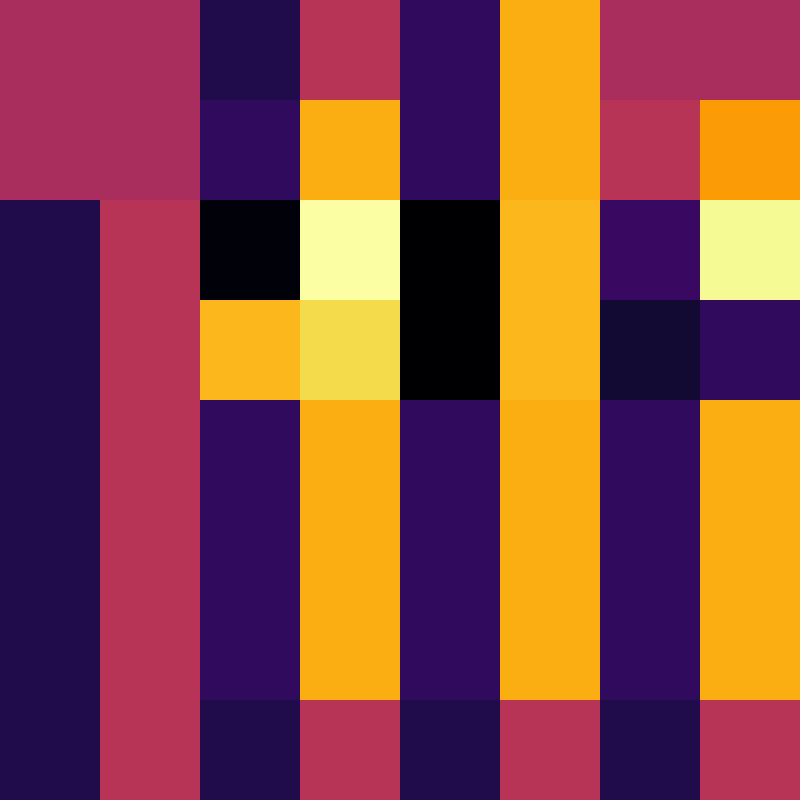} &
\includegraphics[width=0.125\textwidth]{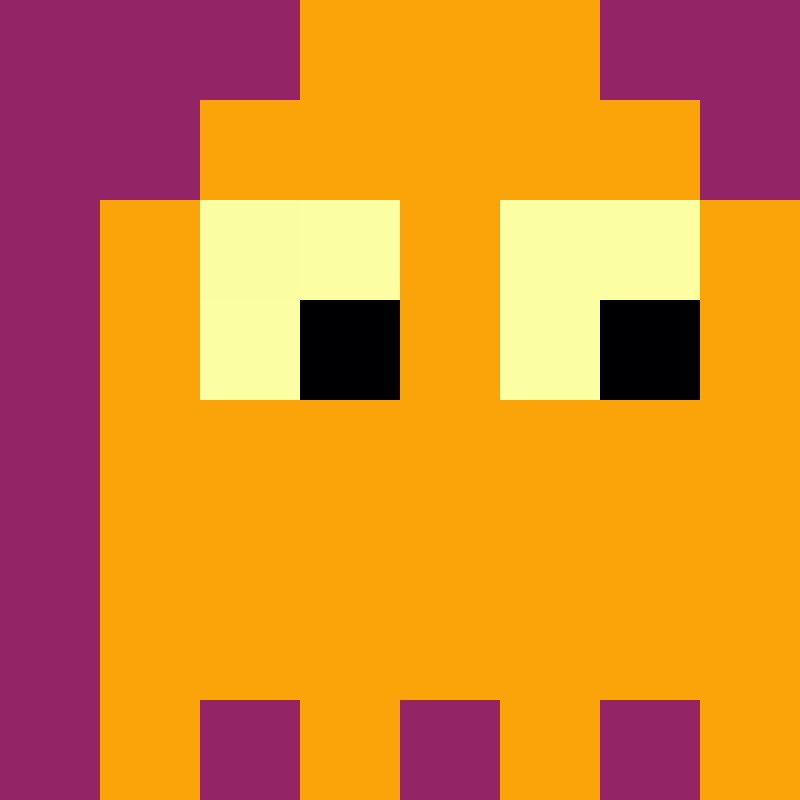} \\
 \small(a) $T=0$ & \small(b) $T=700$ & \small(c) $T=1500$ & \small(d) $T=3000$ & \small(e) $T=5000$ & \small(f) $T=6226$  \\
\end{tabular}
\caption{Sequence of images showing Poincar\'e recurrence in an $8\times 8$ zero-sum polymatrix game, where the changing color of each pixel on the grid represents the mixed strategy of the player over time. After time $T=6226$, we see that an approximation of the original image is recovered, showing that the recurrence property holds.}
\label{fig:clydemain}
\end{figure}

To generate Figure~\ref{fig:clydemain}, we show the data from a simplified 64-player polymatrix game simulation, where the graph that represents player interactions is sparse. Here, the strategy of each player informs the RGB value of a corresponding pixel on a grid. If the system exhibits Poincar\'e recurrence, we should eventually see similar patterns emerge as the pixels change color over time (i.e., as their corresponding mixed strategies evolve). As observed in Figure~\ref{fig:clydemain}, the system returns near the initial image at time $T=6226$. Further details about the experiments can be found in Appendix \ref{app_sec:experiments}.

\section{Discussion}
\label{sec:related}
\label{sec:discussion}
We study both {\gda} and {\ftrl} learning dynamics in periodically varying zero-sum games. We prove that the recurrent nature of such dynamics carries over from static games to the classes of evolving games we study. Yet, in the settings we analyze, the time-average convergence behavior from static zero-sum games can fail to generalize. This work takes a step toward understanding the behavior of classical learning algorithms for games in the more realistic setting where the game itself is not fixed. 

We conclude by discussing related works on learning dynamics in evolving games. The existing literature considers a number of models that admit distinct results~\cite{mai2018cycles, skoulakis2020, cardoso2019competing, duvocelle2018learning}. In a class of time-evolving games where the evolution can be arbitrary~\cite{cardoso2019competing}, algorithms are designed that provide a novel type of regret guarantee called Nash equilibrium regret. In a sense these algorithms are competitive against the Nash equilibrium of the long-term-averaged payoff matrix. In an analysis of discrete-time {\ftrl} dynamics in evolving games that are strictly/strongly monotone~\cite{duvocelle2018learning}, sufficient conditions (e.g., when the evolving game stabilizes) under which the dynamics track/converge to the evolving equilibrium are derived. Unfortunately, zero-sum games do not satisfy these strong properties. Finally, in a model of endogenously evolving zero-sum games where a parametric game evolves itself adversarially towards the participating agents,  a transformation that treats the game as an additional ``hidden" agents allows for a reduction to a more standard static network zero-sum game has been developed~\cite{mai2018cycles, skoulakis2020} under which both time-average convergence to equilibrium as well as Poincar\'{e} recurrence holds. 


This growing literature indicates that time-varying games can exhibit distinct and often times more complex  
behavior than their classic static counterparts. As such, there is much potential for future work towards a better understanding of their learning dynamics.


\section*{Acknowledgements}
 This research/project is supported in part by the National Research Foundation, Singapore under its AI Singapore Program (AISG Award No: AISG2-RP-2020-016), NRF 2018 Fellowship NRF-NRFF2018-07, NRF2019-NRF-ANR095 ALIAS grant, grant PIE-SGP-AI-2018-01, AME Programmatic Fund (Grant No. A20H6b0151) from the Agency for Science, Technology and Research (A*STAR). Tanner Fiez was supported by a National Defense Science and Engineering Graduate Fellowship. 

\typeout{}
\bibliographystyle{abbrvnat}
\bibliography{neurips_refs}

\clearpage

\appendix
\onecolumn

\appendixpage
\addappheadtotoc
\section*{Appendix Organization and Contents}
The organization and contents of this appendix is as follows. Appendix~\ref{app_sec:related} covers additional related work, which we provide a separate bibliography for at the end of the document. Following Appendix~\ref{app_sec:related}, proofs for the theoretical results in the paper are presented in the order that they appeared. Specifically, Appendix~\ref{app_sec:gda_recurr} contains the proofs for the results presented in Section~\ref{sec:gda_recurrent} on {\gda} dynamics and recurrence. This includes the proofs of Proposition~\ref{prop:poincare_failure}, Lemma~\ref{lemma:gdavolume}, Lemma~\ref{lemma:gdabounded}, and Theorem~\ref{thm:gdrec}.
In Appendix~\ref{app_sec:odesolution}, we provide the proof of Proposition~\ref{prop:tagda} from Section~\ref{sec:gda_time_avg} on the time-average behavior of {\gda}.
Appendix~\ref{app_sec:recurrence} contains the analysis for Section~\ref{sec:ftrl_recur} on {\ftrl} dynamics and recurrence, including the proofs of Lemma~\ref{lem:ftrldiv}, Lemma~\ref{lem:boundorbit}, and Theorem~\ref{thm:ftrlrec}. Appendix~\ref{app_sec:taverage} covers the time-average behavior of {\ftrl} dynamics as presented in Section~\ref{sec:ftrltavg}. Specifically, the proofs of Theorem~\ref{thm:tavgconvergence} and Proposition~\ref{prop:tarep} are given in Appendix~\ref{app_sec:taverage}. Finally, Appendix~\ref{app_sec:experiments} contains additional experimental results.

\section{Additional Related Work}
\label{app_sec:related}
The discussions of related work presented in Section~\ref{sec:intro} and Section~\ref{sec:discussion} focused on comparable theoretical results in static zero-sum games and studies on classes of certain evolving zero-sum games. We remark that there is also a rich literature studying evolutionary dynamics in action-dependent (endogenous) evolving games in problems strongly motivated by applications in science, economics, and sociology. Often this line of work falls under the hood of what is known as negative frequency dependent selection~\cite{heino1998enigma}. Negative frequency dependent selection is an evolutionary process in which the fitness of a strategy dissipates as it becomes more common. Indeed, a number of works (see, \cite{weitz2016oscillating, tilman2020evolutionary} and the references therein) analyze evolutionary dynamics in specific formulations, typically in low-dimensional strategy spaces, and characterize the outcomes. In contrast, we study a broad class of learning dynamics in a general class of exogenously evolving games.

\section{{\gda} Recurrence Results: Proofs for Section~\ref{sec:gda_recurrent}}
\label{app_sec:gda_recurr}

This appendix includes the proofs of Proposition~\ref{prop:poincare_failure}, Lemma~\ref{lemma:gdavolume}, Lemma~\ref{lemma:gdabounded}, and Theorem~\ref{thm:gdrec}.

\subsection{Proof of Proposition~\ref{prop:poincare_failure}}
To prove this result, we construct time-evolving zero-sum games without both periodic payoffs and a time-invariant equilibrium in which the {\gda} dynamics are not Poincar\'{e} recurrent.

\textbf{Example 1.} Consider a time-evolving zero-sum game on scalar action spaces so that $x_1, x_2\in \mb{R}$ with the time-evolving payoff matrix $A(t)=t^{-2}$. Without loss of generality, we can consider $t>0$ so that the {\gda} dynamics are well-defined. This time-evolving zero-sum does not have periodic payoffs, but $(x_1^{\ast}, x_2^{\ast})=(0,0)$ is a time-invariant Nash equilibrium. We now show that the {\gda} dynamics are not Poincar\'{e} recurrent in this game. Before formally proving this, we remark that the intuition for why this statement holds is that since the payoff matrix goes to zero,  the distance the dynamics can travel is bounded so it is impossible that the trajectory could return arbitrarily close to an initial condition infinitely often.

The {\gda} dynamics in this time-evolving zero-sum game are described by the system 
\begin{equation*}
\begin{bmatrix}\dot{x}_1\\ \dot{x}_2\end{bmatrix}= \begin{bmatrix}0 & \frac{1}{t^2}\\ -\frac{1}{t^2}& 0\end{bmatrix} \begin{bmatrix}x_1(t)\\ x_2(t)\end{bmatrix}.
\end{equation*}
The solution of a time-varying linear system of this form is given by
\[\begin{bmatrix}x_1(t) \\ x_2(t) \end{bmatrix}=\exp\left(\int_{t_0}^t \begin{bmatrix}0 & \frac{1}{\tau^2}\\ -\frac{1}{\tau^2}& 0\end{bmatrix}d\tau\right)\begin{bmatrix} x_1(t_0) \\ x_2(t_0) \end{bmatrix}.\]
We consider $t_0>1$ without loss of generality. To derive the explicit solution, we begin by computing the integral of the evolving payoff matrix and get that
\[\int_{t_0}^t \begin{bmatrix}0 & \frac{1}{\tau^2}\\ -\frac{1}{\tau^2}& 0\end{bmatrix}d\tau = \begin{bmatrix} 0 & \frac{1}{t_0}-\frac{1}{t} \\ \frac{1}{t}-\frac{1}{t_0}&0 \end{bmatrix}.\]
Recalling the following identity
\[\exp\left(\theta \bmat{0 & -1\\ 1 & 0}\right)=\bmat{\cos(\theta) & \sin(\theta)\\ -\sin(\theta) & \cos(\theta)},\]
we can determine that the matrix exponential is then given by
\[\exp\left(\begin{bmatrix} 0 & \frac{1}{t_0}-\frac{1}{t} \\ \frac{1}{t}-\frac{1}{t_0}&0 \end{bmatrix}\right)=\begin{bmatrix} \cos(\frac{1}{t}-\frac{1}{t_0}) & -\sin(\frac{1}{t}-\frac{1}{t_0}) \\ \sin(\frac{1}{t}-\frac{1}{t_0}) & \cos(\frac{1}{t}-\frac{1}{t_0}) \end{bmatrix}.\]
Therefore, the solution of the system simplifies to be given by
\[\begin{bmatrix} x_1(t)\\ x_2(t)\end{bmatrix}=\begin{bmatrix} \cos(\frac{1}{t}-\frac{1}{t_0}) & -\sin(\frac{1}{t}-\frac{1}{t_0}) \\ \sin(\frac{1}{t}-\frac{1}{t_0}) & \cos(\frac{1}{t}-\frac{1}{t_0}) \end{bmatrix}\begin{bmatrix} x_1(t_0)\\ x_2(t_0)\end{bmatrix},\]
and equivalently, 
\[\begin{bmatrix} x_1(t)\\ x_2(t)\end{bmatrix}=\begin{bmatrix} \cos(\frac{1}{t}-\frac{1}{t_0})x_1(t_0)  -\sin(\frac{1}{t}-\frac{1}{t_0})x_2(t_0) \\ \sin(\frac{1}{t}-\frac{1}{t_0})x_1(t_0) + \cos(\frac{1}{t}-\frac{1}{t_0})x_2(t_0) \end{bmatrix}.\]

Given the explicit form of the solution, we now show that we can construct an initial condition that the strategies do not return back to infinitely often. This will immediately allow us to conclude the system is not Poincar\'{e} recurrent by definition. We remark that the following choice of initial condition is only for the simplicity of the proof and identical conclusions would hold for almost all initial conditions.

Let $x_1(t_0)=1$ and $x_2(t_0)=0$. Given this initial condition, the solution of the system simplifies to be given by
\[\begin{bmatrix} x_1(t)\\ x_2(t)\end{bmatrix}=\begin{bmatrix} \cos(\frac{1}{t}-\frac{1}{t_0})\\ \sin(\frac{1}{t}-\frac{1}{t_0}) \end{bmatrix}.\]
Taking the limit of the solution as $t\rightarrow \infty$, we have that 
\[\lim_{t\rightarrow\infty}x_1(t)=\lim_{t\rightarrow\infty}\cos\Big(\frac{1}{t}-\frac{1}{t_0}\Big)=\cos\Big(\frac{1}{t_0}\Big)\]
and 
\[\lim_{t\rightarrow\infty}x_2(t)= \lim_{t\rightarrow\infty}\sin\Big(\frac{1}{t}-\frac{1}{t_0}\Big)=-\sin\Big(\frac{1}{t_0}\Big).\]
This shows that the dynamics converge to a fixed point $(\bar{x}_1,\bar{x}_2)=(\cos\big(\tfrac{1}{t_0}\big), -\sin\big(\tfrac{1}{t_0}\big))$. Since $(\bar{x}_1,\bar{x}_2)=(\cos\big(\tfrac{1}{t_0}\big), \sin\big(\tfrac{1}{t_0}\big))\neq (1, 0)=(x_1(t_0), x_2(t_0))$ for $t_0>1$ unless $t_0\rightarrow\infty$,
this immediately implies that the {\gda} dynamics are not Poincar\'{e} recurrent in this time-evolving zero-sum game since they do not return infinitely often back to an arbitrarily small neighborhood around the initial condition. 

\textbf{Example 2.} In the previous example, we showed that the {\gda} dynamics were not Poincar\'{e} recurrent in a time-evolving zero-sum game that had a time-invariant Nash equilibrium but not periodic payoffs. In this example, we provide theoretical evidence that the {\gda} dynamics are not Poincar\'{e} recurrent in a time-evolving zero-sum game with periodic payoffs but without a time-invariant Nash equilibrium.

Consider a time-evolving zero-sum game on scalar action spaces so that $x_1, x_2\in \mb{R}$ with a periodic payoff matrix $A(t)=A(t+T)$ for any $t\geq 0$ and $T=3$ that evolves over a period such that $A(t)=1$ for $0\leq t\leq 1$ and $A(t)=-1$ for $1\leq t\leq 3$. We treat player 2 as a dummy player that just plays the fixed strategy of $x_2=1$ for all $t$. Thus, this time-evolving zero-sum game can be viewed as a trivial game that is equivalent to an optimization problem for player 1. Since player 1 is a utility maximizer, the Nash equilibrium of the game at each time simply corresponds to the strategy of player 1 that maximizes its utility. Thus, the Nash equilibrium when $A(t)=1$ is $x_1^{\ast}= \infty$ and given $A(t)=-1$ it is $x_21^{\ast}= -\infty$. Therefore, this corresponds to a time-evolving zero-sum game that is periodic but there is not a time-invariant Nash equilibrium.

We now show that the {\gda} dynamics are not Poincar\'{e} recurrent in this game. The dynamics can 
be described by the system $\dot{x}_1=A(t)$. Consequently, the solution is $x_1(t)=x_1(t_0)+t$ on the interval with $A(t)=1$ and $x_1(t)=x_1(t_0)-1$ on the interval with $A(t)=-1$. This means that after 1 period of the game, $x_1(t)= x_1(t_0)-1$, which implies that $x_1(t)\rightarrow -\infty$ as $t\rightarrow\infty$. Thus the dynamics are not Poincar\'{e} recurrent in this time-evolving zero-sum game since they do not return infinitely often back to an arbitrarily small neighborhood around the initial condition. 

\begin{figure}
    \centering
    \includegraphics[width=\textwidth]{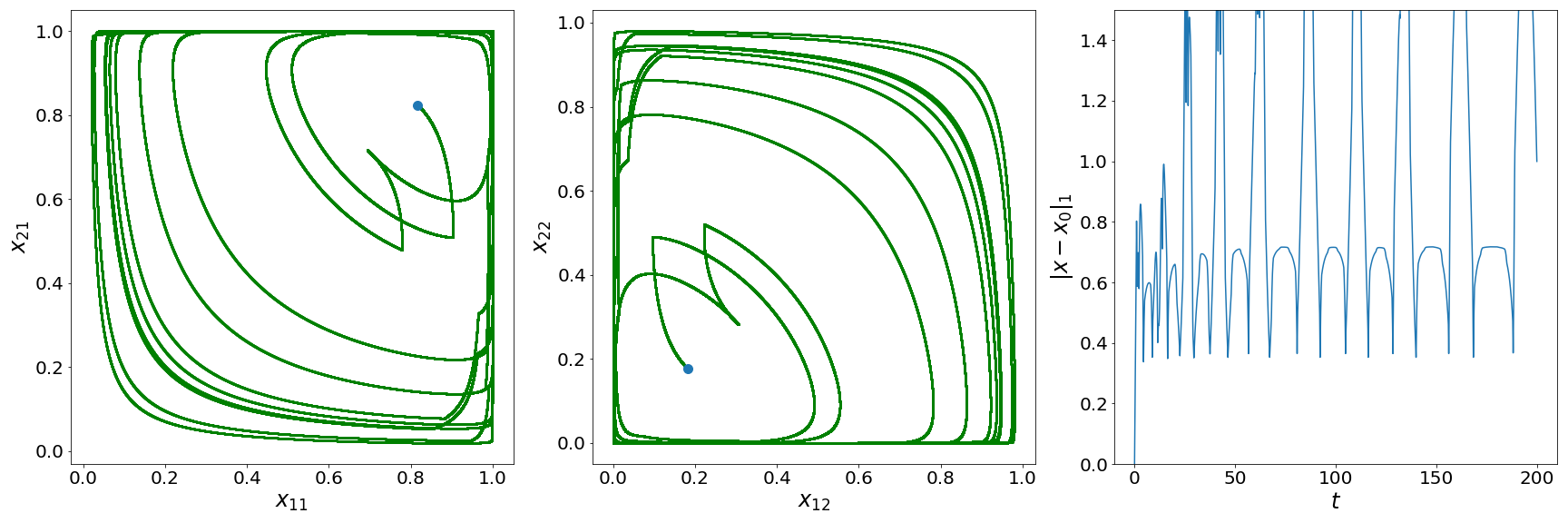}
    \caption{(Left, Center) Replicator trajectories for periodically evolving game without time-invariant equilibrium. (Right) $L_1$-norm plot showing that recurrence does not hold in this example.}
    \label{fig:counterexample_prop1}
\end{figure}

\textbf{Example 3.}
Additionally, we provide an experimental example to show the non-existence of Poincar\'{e} recurrence in the setting of \ftrl~dynamics. In particular, we simulate a time-evolving zero-sum game which has periodic payoffs but does not have a time-invariant Nash equilibrium.
Consider a time-evolving zero-sum game where the payoff matrix for the first quarter of a period is standard Matching Pennies $A =  \bigl(\begin{smallmatrix}1 & -1\\ -1 & 1\end{smallmatrix} \bigr)$. For the remaining three quarters of the period it is instead $A =   \bigl(\begin{smallmatrix}0.05 & -0.5\\ -0.5 & 5\end{smallmatrix} \bigr)$. Note that here the payoff matrix for the second player is just $-A$. The former game has a mixed Nash equilibrium where both players play $[0.5, 0.5]$ and the latter game has a mixed Nash equilibrium where both players play $[0.9091, 0.0909]$. We simulate this example with replicator dynamics, which is an instantiation of \ftrl~dynamics. First, we plot the trajectories of the player’s strategies against each other. Moreover, we plot the $L_1$-norm between the joint trajectory of the players and the initial condition. The simulation results show that the trajectory does not return back arbitrarily close to the initial condition, thus the dynamics are not Poincar\'{e} recurrent in this periodic evolving game without a time-invariant equilibrium (Figure \ref{fig:counterexample_prop1}).

\subsection{Proof of Lemma~\ref{lemma:gdavolume}}
We can show that the {\gda} dynamics are volume preserving by showing that the vector field is divergence free and then applying Liouville's theorem. Indeed,
\[\text{div}(\dot{x})=\sum_{i=1}^2\sum_{j=1}^{n_i}\frac{\partial \dot{x}_{ij}}{\partial x_{ij}}=0,\]
which follows from the fact that $\dot{x}_{ij}$ is independent of $x_{ij}$ for each $i, j$.
The divergence free property of the vector field then ensures 
that the flow $\phi^t$ of the differential equation is volume preserving by Liouville's theorem.

\subsection{Proof of Lemma~\ref{lemma:gdabounded}}
To prove this statement, we claim the following function is time-invariant:
\[\Phi(t)=\frac{1}{2}\big(x_1^\top(t) x_1(t)+x_2^\top(t) x_2(t)\big).\]
By taking the time-derivative of the $\Phi(t)$ we can verify the function is a constant of motion. Indeed, this holds based on the following analysis:
\begin{align*}
\frac{d\Phi}{dt}&= \frac{1}{2}\big(x_1^\top(t)\dot{x}_1+\dot{x}_1^\top x_1(t)+x_2^\top(t) \dot{x}_2+\dot{x}_2^\top x_2(t)\big)\\
&= x_1^\top(t)\dot{x}_1+x_2^\top(t) \dot{x}_2\\
&= x_1^\top(t) A(t)x_2(t)-x_2(t)^\top A^\top(t) x_1(t) \\
&=0.
\end{align*}
Finally, observe that given a bounded initial condition, the time-invariance of $\Phi(t)$ directly implies that no strategy of any player can become unbounded so the flow $\phi^t$ of the differential equation has bounded orbits.

\subsection{Proof of Theorem~\ref{thm:gdrec}}
Given the previous intermediate results, Theorem~\ref{thm:gdrec} follows from the arguments presented in Section~\ref{sec:periodic}. In particular, 
observe that by definition of the periodic zero-sum bilinear game, the {\gda} dynamics are $T$-periodic. Now, consider the discrete-time dynamical system defined by the Poincar\'{e} map $\phi^{T}$ that arises. This system retains the volume preservation property of the continuous-time system from Lemma~\ref{lemma:gdavolume} since as presented in Section~\ref{sec:periodic}, if a $T$-periodic system is divergence-free then the discrete-time system defined by $\phi^{T}$ is also volume preserving~\cite[3.16.B, Thm 2]{arnold2013mathematical}.
Similarly, the discrete-time system defined by the $\phi^{T}$ retains the bounded orbits guarantee of the continuous-time system from Lemma~\ref{lemma:gdabounded} since it holds at any set of times. Thus, we are able to apply the  Poincar\'{e} recurrence theorem for discrete-time systems from Section~\ref{sec:periodic} to the discrete-time system defined by $\phi^T$ to conclude the discrete-time system is Poincar\'{e} recurrent. This immediately implies that the {\gda} dynamics are Poincar\'{e} recurrent since the discrete-time system defined by $\phi^T$ forms a subsequence of the continuous-time system.

\section{{\gda} Time-Average Result: Proof of Proposition~\ref{prop:tagda}}
\label{app_sec:odesolution}
Consider a periodic zero-sum bilinear game with $x_1,x_2\in \mb{R}$ and a periodic payoff matrix $A(t)$ such that $A(t)=A(t+T)$ with $T=3\pi$ for any $t\geq 0$. 
Moreover, let the payoff matrix evolve over a period as follows:
 \[ A(t) = \left\{
\begin{array}{ll}
      -1 & 0\leq t \leq \pi \\
      \ \ \ 1 & \pi \leq t\leq \frac{3\pi}{2}\\
      -1 &  \frac{3\pi}{2}\leq t\leq 3\pi.
\end{array} 
\right. \]
The joint strategy $(x_1^{\ast}, x_2^{\ast})=(0,0)$ is the time-invariant Nash equilibrium. We now show that the time-average of the strategies produced by the {\gda} dynamics do not converge to the time-invariant Nash equilibrium.


The {\gda} dynamics in this periodic zero-sum bilinear game are given by
\begin{align*}
\dot{x}_1&= A(t)x_2(t)\\
\dot{x}_2 &= -A(t)x_1(t).
\end{align*}
The solution to the differential equation that describes the {\gda} dynamics can be constructed in a piecewise manner. On each of the three intervals we have a linear system defined by
\begin{equation*}
\begin{bmatrix}\dot{x}_1\\ \dot{x}_2\end{bmatrix}= \begin{bmatrix}0 & A(t)\\ -A(t) & 0\end{bmatrix} \begin{bmatrix}x_1\\ x_2\end{bmatrix}.
\end{equation*}
Now recall the following identity
\[\exp\left(\theta \bmat{0 & 1\\ -1 & 0}\right)=\bmat{\cos(\theta) & \sin(\theta)\\ -\sin(\theta) & \cos(\theta)}.\]
Hence for initial condition $(x_1(0),x_2(0))$ and interval $[0,\pi)$ we know that $A(t)=-1$ for all $t$ in the interval which implies that the solution on this interval is given by
\[\bmat{x_1(t)\\ x_2(t)}=\bmat{\cos(-t) & \sin(-t)\\ -\sin(-t) & \cos(-t)}\bmat{x_1(0)\\ x_2(0)}.\]
On the interval $[\pi,3\pi/2)$, $A(t)=1$ so that
\[\bmat{x_1(t)\\ x_2(t)}=\bmat{\cos(t-\pi) & \sin(t-\pi)\\ -\sin(t-\pi) & \cos(t-\pi)}\bmat{x_1(\pi)\\ x_2(\pi)}.\]
Finally, on $[3\pi/2,3\pi)$, $A(t)=-1$ so that
\[\bmat{x_1(t)\\ x_2(t)}=\bmat{\cos(-(t-3\pi/2)) & \sin(-(t-3\pi/2))\\ -\sin(-(t-3\pi/2)) & \cos(-(t-3\pi/2))}\bmat{x_1(3\pi/2)\\ x_2(3\pi/2)}.\]
Now, let us consider the initial condition $(x_1(0),x_2(0))=(1,0)$. Then, 
\[(x_1(t),x_2(t))=\left\{\begin{array}{ll}
    (\cos(t),\sin(t))=(\cos(-t),-\sin(-t)) & \ t\in[0,\pi)  \\
    (\cos(t),-\sin(t))=(-\cos(t-\pi),\sin(t-\pi)) &t\in[\pi,3\pi/2)\\
    (-\cos(t),-\sin(t))=(\sin(-(t-3\pi/2)), \cos(-(t-3\pi/2)))) & \ t\in[3\pi/2,3\pi)\end{array}\right.\]
Observe that from this solution we can determine that the {\gda} dynamics return to the initial condition 
at the end of a period. Thus, to assess convergence of the time-average it is sufficient to evaluate the time-average of the dynamics over a period of the evolving game. Integrating the solution over a period, we have that
\begin{align*}
\int_0^{3\pi}x_1(t)dt&=\int_0^{3\pi/2}\cos(t)dt+\int_{3\pi/2}^{3\pi}-\cos(t)dt\\
&=[\sin(3\pi/2)-\sin(0)]-[\sin(3\pi)-\sin(3\pi/2)] \\
&=-2
\end{align*}
and
\begin{align*}
\int_0^{3\pi}x_2(t)dt&=\int_0^{\pi}\sin(t)dt+\int_{\pi}^{3\pi}-\sin(t)dt\\
&=[-\cos(\pi)+\cos(0)]+[\cos(3\pi)-\cos(\pi)] \\
&=2
\end{align*}
This implies that the time-average strategies of the players do not equal to zero, so the time-average of the {\gda} dynamics do not converge to the time-invariant Nash equilibrium. The fact that it is non-zero holds generally even changing the initial condition. 

This completes the proof and shows that there exists periodic zero-sum bilinear games where the time-average {\gda} strategies do not converge to the time-invariant Nash equilibrium.

\section{{\ftrl} Poincar\'{e} Recurrence Results: Proofs for Section~\ref{sec:ftrl_recur}}
\label{app_sec:recurrence}
This appendix includes the proofs of Lemma~\ref{lem:ftrldiv}, Lemma~\ref{lem:boundorbit}, and Theorem~\ref{thm:ftrlrec}.
\subsection{Proof of Lemma~\ref{lem:ftrldiv}}
\label{app_sec:ftrldivproof}
Recall that this result states that the dynamics defined by the system $\dot{z}$ (given again in~\eqref{eq:zdot}) are volume preserving in any periodic zero-sum polymatrix game. Here, we also explain how the $\dot{z}$ dynamics were formulated. Then, we show that the divergence of this vector field is zero, from which we conclude the dynamics are volume preserving by Liouville's theorem. This proof follows closely arguments in~\cite{mertikopoulos2018cycles}.

For each player $i \in V$, given a fixed strategy $\beta \in \A_i$, for all $\alpha\in \A_i\setminus \beta$ the cumulative utility differences are defined by
\[z_{i\alpha}(t)=y_{i\alpha}(t)-y_{i\beta}(t).\]
This transformation from the cumulative utilities to the cumulative utility differences yields a linear map $\Pi_i:\mb{R}^{n_i}\rightarrow \mb{R}^{n_i-1}$ from $y_i(t)$ to $z_i(t)$ for each player $i \in V$. Moreover, define by $\Pi=(\Pi_1,\dots, \Pi_{|V|})$ the product map of the linear maps $\Pi_i$ of each player $i\in V$. This map is surjective, but not injective.

Observe that the cumulative utility differences for each player $i \in V$ and all $\alpha\in \A_i\setminus \beta$ evolve following the differential equation
\begin{equation}
\dot{z}_{i\alpha}(t)=v_{i\alpha}(x(t),t)-v_{i\beta}(x(t),t).
\label{eq:zdot}
\end{equation}
The above differential equation is obtained directly by the form of $y_i$ and the fundamental theorem of calculus. Moreover, recall that for any player $i \in V$ and pure strategy $\gamma \in \A_i$, the quantity $v_{i\gamma}(x(t),t)$ gives utility of player $i \in V$ at any time $t\geq 0$ for selecting the pure strategy $\gamma \in \A_i$. 

To analyze the dynamics from the system in~\eqref{eq:zdot} we need it to be well-defined, which is not immediate since it depends on $x(t)=Q(y(t))$ and the mapping from $y(t)$ to $z(t)$ via $\Pi$ is not invertible so that $y(t)$ cannot be expressed as a fuction of $z(t)$. However, despite this, the system is in fact well-defined. To see this, for each player $i\in V$, consider the reduced choice map $\hat{Q}_i:\mb{R}^{n_i-1}\rightarrow \X_i$ defined as $\hat{Q}_i(z_i(t))= Q_i(y_i(t))$
for some $y_i(t)\in \mb{R}^{n_i}$ such that $\Pi_i(y_i(t))=z_i(t)$ which is guaranteed to exist since $\Pi_i$ is surjective. Then, the fact that $\hat{Q}_i(z_i(t))$ is well-defined for each player $i\in V$ holds since by the construction $\Pi_i(y_i(t))=\Pi_i(y_i'(t))$ if and only if $y_{i\alpha}'(t)=y_{i\alpha}(t)+c$ for $c\in \mb{R}$ and every $\alpha_i\in \A_i$ which immediately implies $Q_i(y_i'(t))=Q_i(y_i(t))$ if and only if $\Pi_i(y_i(t))=\Pi_i(y_i'(t))$. Finally, let 
$\hat{Q}=(\hat{Q}_1,\dots, \hat{Q}_{|V|})$ be the combined reduced choice map and note that $Q(y(t))=\hat{Q}(\Pi(y(t))=\hat{Q}(z(t))$ by the construction.
As a result, the dynamics from the system in~\eqref{eq:zdot} are equivalently given by the following system
\[\dot{z}_{i\alpha}=v_{i\alpha_i}(\hat{Q}_i(z(t), t))-v_{i\beta_i}(\hat{Q}_i(z(t)), t).\]
This system is well-defined by the arguments above which ensures that the system in~\eqref{eq:zdot} is well-defined.

Now that we have shown the system is well-defined, we prove that is is volume preserving. To see this, observe that the vector field is divergence free. Indeed,
\[\text{div}(\dot{z})=\sum_{i\in V}\sum_{\alpha\in \A_i}\frac{\partial \dot{z}_{i\alpha}}{\partial z_{i\alpha}}=\sum_{i\in V}\sum_{\alpha\in \A_i}\sum_{\gamma_i\in \A_i}\frac{\partial \dot{z}_{i\alpha}}{\partial x_{i\gamma}}\frac{\partial x_{i\gamma}}{\partial z_{i\alpha_i}}=0.\]
Note that the equation above holds since for each player $i\in V$, the pure strategy utilities at any time $t\geq 0$ given by $v_i(x(t), t)$ where $v_{i\alpha}(x(t), t)=u_i((\alpha, x_{-i}(t)), t)$  do not depend on $x_i(t)$. 
Finally, the divergence free property of the vector field ensures 
that the flow $\phi^t$ of the differential equation is volume preserving by Liouville's theorem.

\subsection{Proof of Lemma~\ref{lem:boundorbit}}
\label{app_sec:ftrlboundedproof}
Recall that Lemma~\ref{lem:boundorbit} states that the orbits of the $\dot{z}$ dynamics are bounded. To prove this statement, we show that the function 
\[\Phi(x^{\ast}, y(t))=\sum_{i\in V}\big(h_i^{\ast}(y_i(t)) - \langle x_i^{\ast},y_i(t) \rangle + h_i(x_i^{\ast})\big)\]
is time-invariant where $x^{\ast}$ denotes the time-invariant fully mixed Nash equilibrium and then argue that this is sufficient to ensure that orbits of the $\dot{z}$ dynamics are bounded.

To prove that the function $\Phi(x^{\ast}, y(t))$ is time-invariant, we show that the time-derivative of the function is equal to zero. The time-derivative of $\Phi(x^{\ast}, y(t))$ simplifies using the fact that $h_i(x_i^{\ast})$ is time-independent to the following:
\begin{align*}
\frac{d\Phi(x^{\ast}, y(t))}{dt}
&= \frac{d}{dt}\sum_{i\in V} h_i^{\ast}(y_i(t)) + \frac{d}{dt}\sum_{i\in V}\langle x_i^{\ast}, y_i(t) \rangle.
\end{align*}
We begin by showing that the time-derivative of $\sum_{i\in V} h_i^{\ast}(y_i(t))=0$. This holds by the following computation that is explained below:
\begin{align}
\frac{d}{dt}\sum_{i\in V} h_i^{\ast}(y_i(t))
&= \sum_{i\in V} \langle \nabla h_i^{\ast}(y_i(t)), \dot{y}_i(t) \rangle \label{eq:chain}\\
&= \sum_{i\in V} \langle  x_i(t), \dot{y}_i(t) \rangle \label{eq:maximizing}  \\
&= \sum_{i\in V} \langle x_i(t),  v_i(x(t),t) \rangle \label{eq:fundamental}\\
&= \sum_{i\in V} u_i(x(t), t) \label{eq:def}\\
&=0.\label{eq:zerosum}
\end{align}
We obtain~\eqref{eq:chain} by the chain rule,~\eqref{eq:maximizing} by the maximizing argument of convex conjugates (see e.g.,~\citealt[Chapter~2]{shalev2011online}) that implies $x_i(t)=Q_i(y_i(t))= \nabla h_i^\ast(y_i(t))$,~\eqref{eq:fundamental} by the definition of $y_i(t)$ and the fundamental theorem of calculus,~\eqref{eq:def} by definition of the pure strategy utilities $v_i(x(t),t)$ and the utility $u_i(x(t),t)$, and~\eqref{eq:zerosum} by the fact that the polymatrix game is zero-sum.


We now finish by showing that the time-derivative of $\sum_{i\in V}\langle x_i^{\ast}, y_i(t) \rangle=0$. To begin, observe that the time-derivative can be described by
\begin{align}
\frac{d}{dt}\sum_{i\in V}\langle x_i^{\ast}, y_i(t) \rangle&=\sum_{i\in V}\langle x_i^{\ast}, \dot{y}_i(t) \rangle \notag \\
&= \sum_{i\in V} \sum_{j:(i,j) \in E}(x_i^{\ast})^{\top}A^{ij}(t)x_j(t) \label{eq:def2}\\
&=\sum_{i\in V} \sum_{j:(i,j) \in E}(x_i^{\ast})^{\top}A^{ij}(t)(x_j(t)-x_j^{\ast}). \label{eq:minusu}
\end{align}
Observe that~\eqref{eq:def2} follows from the definition of $y_i(t)$ and the fundamental theorem of calculus and~\eqref{eq:minusu} comes about from subtracting $\sum_{i\in V}u_i(x^{\ast}, t)$ which is zero by the fact that the polymatrix game is zero-sum for any $t\geq 0$.

To continue, we remark that any zero-sum polymatrix game can be transformed to a payoff equivalent, pairwise constant-sum game~\cite{cai2011minmax}. 
This means that for each edge $(i,j) \in E$ there exists a matrix $B^{ij}(t)$ such that the following properties hold (see Lemma~3.1, 3.2, and 3.4, respectively \citealt{cai2011minmax}):
\begin{description}
    \item[Property 1.] $ A^{ij}_{\alpha \beta}(t) -  A^{ij}_{\alpha \gamma}(t) = B^{ij}_{\alpha \beta}(t) - B^{ij}_{\alpha \gamma}(t)$ for any pure strategies $\alpha \in \A_i$ and $\beta,\gamma \in \A_j$.
    \item[Property 2.] $B^{ij}(t) + (B^{ji}(t))^\top = c_{ij}(t) \cdot \onev_{n_i \times n_j}$, where $c_{ij}(t)$ is a constant and $\onev_{n_i \times n_j}$ is an $n_i\times n_j$ matrix of ones. 
    \item[Property 3.] In every joint pure strategy profile, every player $i \in V$ has the same utility in the game defined by the individual payoff matrices $\{ A^{ij}(t)\}_{(i,j)\in E}$ as in the game defined by the individual payoff matrices $\{B^{ij}(t)\}_{(i,j)\in E}$. 
\end{description}
\smallskip
\noindent Fixing a strategy $\gamma \in \A_j$, we can equivalently express any summand of~\eqref{eq:minusu} in the following manner that is justified below:
\begin{align}
(x_i^\ast)^\top   A^{ij}  (x_j(t)-x_j^\ast) &= 
\sum_{\alpha \in \A_i}\sum_{\beta \in \A_j}
x_{i\alpha}^\ast  A^{ij}_{\alpha \beta}(x_{j\beta}(t)-x_{j\beta}^\ast) \notag \\
&\mkern-44mu= \sum_{\alpha \in \A_i}\sum_{\beta \in \A_j}
x_{i\alpha}^\ast \big( B^{ij}_{\alpha \beta}(t) - B^{ij}_{\alpha \gamma}(t) +   A^{ij}_{\alpha \gamma}(t) \big) (x_{j_\beta}(t)-x_{j_\beta}^\ast) \label{eq:prop1} \\
&\mkern-44mu= (x_i^\ast)^\top  B^{ij}(t)  (x_j(t)-x_j^\ast) + \sum_{\alpha \in \A_i} x_{i\alpha}^{\ast}\left( A^{ij}_{\alpha \gamma}(t)-B^{ij}_{\alpha \gamma}(t) \right) \sum_{\beta \in \A_j}(x_{j\beta}(t)-x_{j\beta}^\ast). 
\notag \\
&\mkern-44mu=(x_i^\ast)^\top  B^{ij}(t)  (x_j(t)-x_j^\ast).\label{eq:zerosimple} 
\end{align}
Observe that~\eqref{eq:prop1} results from applying Property~1 and~\eqref{eq:zerosimple} holds since both $x_j(t)$ and $x_j^{\ast}$ are on the simplex so that $\sum_{\beta \in \A_j}x_{j\beta}=1$ and $\sum_{\beta \in \A_j}x_{j\beta}^{\ast}=1$ which implies $\sum_{\beta \in \A_j}(x_{j\beta}-x_{j\beta}^\ast)=0$.

Thus, continuing from~\eqref{eq:minusu} and using that $(x_i^\ast)^\top   A^{ij}  (x_j(t)-x_j^\ast)=(x_i^\ast)^\top  B^{ij}(t)  (x_j(t)-x_j^\ast)$ from above and swapping the sum indexing and taking the transpose of the quadratic form $(x_i^\ast)^\top B^{ij}(t)(x_j(t)-x_j^\ast)$, we get that
\begin{align}
\frac{d}{dt}\sum_{i\in V}\langle x_i^{\ast}, y_i(t) \rangle 
&=\sum_{i\in V} \sum_{j:(i,j) \in E}(x_i^{\ast})^{\top}A^{ij}(t)(x_j(t)-x_j^{\ast})\notag \\
&=\sum_{i\in V} \sum_{j:(i,j) \in E}(x_i^{\ast})^{\top}B^{ij}(t)(x_j(t)-x_j^{\ast})\\
&= \sum_{j \in V}\sum_{i:(j, i)\in E}(x_j(t) - x_j^\ast)^\top  (B^{ij}(t))^\top x_i^\ast. \notag 
\end{align}
Moreover, we obtain the following expression that is justified below:
\begin{align}
\frac{d}{dt}\sum_{i\in V}\langle x_i^{\ast}, y_i(t) \rangle 
&= \sum_{j \in V}\sum_{i:(j, i)\in E}(x_j(t) - x_j^\ast)^\top  (B^{ij}(t))^\top x_i^\ast \notag \\
&\mkern-44mu= \sum_{j \in V}\sum_{i:(j, i)\in E}(x_j(t) - x_j^\ast)^\top  (c^{ji}(t)  \onev_{n_j \times n_i}  - B^{ji}(t)) x_i^\ast \label{eq:prop2} \\
&\mkern-44mu= \sum_{j \in V}\sum_{i:(j, i)\in E}c^{ji}(t)(x_j(t) - x_j^\ast)^\top   \onev_{n_j \times n_i}x_i^\ast  - \sum_{j \in V}\sum_{i:(j, i)\in E}(x_j(t) - x_j^\ast)^\top B^{ji}(t)x_i^\ast  \notag \\
&\mkern-44mu= - \sum_{j \in V}\sum_{i:(j, i)\in E}(x_j(t) - x_j^\ast)^\top B^{ji}(t)x_i^\ast. \label{eq:dropc}
\end{align}
Note that~\eqref{eq:prop2} results from applying Property~2 and we obtain~\eqref{eq:dropc}
using that $\sum_{j \in V}\sum_{i:(j, i)\in E}c^{ji}(t)(x_j(t) - x_j^\ast)^\top=0$ since each summand is zero as can be seen by noting that $\sum_{\alpha\in \A_j}x_{j\alpha} = \sum_{\alpha\in \A_j}x_{j\alpha}^{\ast} = \sum_{\alpha\in \A_i}x_{i\alpha}^{\ast} =1$ which gives
\begin{equation*}
c^{ji}(t) (x_j(t) - x_j^\ast)^\top \onev_{n_j \times n_i}  x_i^\ast = c^{ji} (x_j(t) - x_j^\ast)^\top \onev_{n_j} = c^{ji}(t)-c^{ji}(t)  = 0.
\end{equation*}

We now analyze the summand in \eqref{eq:dropc} for some $j \in V$. Fixing any pure strategy $\gamma_i\in \A_i$ for each $i\in V\setminus \{j\}$, obtain the following simplification that is explained below:
\begin{align}
 \sum_{i:(j,i)\in E}(x_j(t)-x_j^\ast)^\top B^{ji}(t)x_i^\ast&=\sum_{i:(j,i)\in E}\sum_{\alpha\in \A_j}\sum_{\beta\in \A_i}  (x_{j\alpha}(t)-x_{j\alpha}^\ast) B^{ji}_{\alpha\beta}(t)x_{i\beta}^\ast \notag \\
&\mkern-244mu=\sum_{i:(j,i)\in E}\sum_{\alpha\in \A_j}\sum_{\beta\in \A_i} (x_{j\alpha}(t)-x_{j\alpha}^\ast) \big(A^{ji}_{\alpha\beta}(t)-A^{ji}_{\alpha\gamma_i}(t)+B^{ji}_{\alpha\gamma_i}(t)\big)x_{i\beta}^\ast \label{eq:prop1second} \\
&\mkern-244mu=\sum_{i:(j,i)\in E}(x_j(t)-x_j^\ast)^\top A^{ji}(t)x_i^\ast  +\sum_{\alpha\in \A_j}(x_{j\alpha}(t)-x_{j\alpha}^\ast) \sum_{i:(j,i)\in E}(B_{\alpha\gamma_i}^{ji}(t)-A^{ji}_{\alpha\gamma_i}(t))\sum_{\beta\in\A_i}x_{i\beta}^\ast
\notag \\
&\mkern-244mu=\sum_{i:(j,i)\in E}(x_j(t)-x_j^\ast)^\top A^{ji}(t)x_i^\ast  +\sum_{\alpha\in \A_j}(x_{j\alpha}(t)-x_{j\alpha}^\ast) \sum_{i:(j,i)\in E}(B_{\alpha\gamma_i}^{ji}(t)-A^{ji}_{\alpha\gamma_i}(t))
\label{eq:simp1} \\
&\mkern-244mu=\sum_{i:(j,i)\in E}(x_j(t)-x_j^\ast)^\top A^{ji}(t)x_i^\ast  
\label{eq:simp2}.
\end{align}
The equation in~\eqref{eq:prop1second} follows from applying Property~1 and the equation in~\eqref{eq:simp1} holds since $\sum_{\beta\in\A_i}x_{i\beta}^\ast=1$ as a result of the strategy spaces being on the simplex. Finally, to see how~\eqref{eq:simp2} is obtained, observe that for each $\alpha \in \A_j$ the terms $\sum_{i:(j,i)\in E}A^{ji}_{\alpha\gamma_i}(t)$ and $\sum_{i:(j,i)\in E}B^{ji}_{\alpha\gamma_i}(t)$ give the utility of player $j \in V$ in the games with payoffs $\{ A^{ji}(t)\}_{(j,i)\in E}$ and $\{B^{ji}(t)\}_{(j,i)\in E}$ respectively under a joint pure strategy. Hence, by Property 3, the respective utilities are equal so that the difference is zero.

Finally, relating~\eqref{eq:simp2} back to~\eqref{eq:dropc}, we conclude that the time-derivative is zero:  
\begin{align*}
\frac{d}{dt}\sum_{i\in V}\langle x_i^{\ast}, y_i(t) \rangle 
&= - \sum_{j \in V}\sum_{i:(j, i)\in E}(x_j(t) - x_j^\ast)^\top B^{ji}(t)x_i^\ast\\
&= - \sum_{j \in V}\sum_{i:(j, i)\in E}(x_j(t) - x_j^\ast)^\top A^{ji}(t)x_i^\ast=0.
\end{align*}
The final equality holds since $x^\ast$ is an interior Nash equilibrium, which implies  $u_{j\alpha}(x^\ast,t)=u_j(x^\ast,t)$ for all strategies $\alpha\in \A_j$ and any linear combination thereof.

Hence,
\begin{align*}
\frac{d\Phi(x^{\ast}, y(t))}{dt}
&= \frac{d}{dt}\sum_{i\in V} h_i^{\ast}(y_i(t)) + \frac{d}{dt}\sum_{i\in V}\langle x_i^{\ast}, y_i(t) \rangle=0,
\end{align*}
which implies that $\Phi(x^{\ast}, y(t))$ is time-invariant.

Finally, by Lemma D.2 of~\citet{mertikopoulos2018cycles}, the time-invariance of $\Phi$ is sufficient to ensure that the flow $\phi^t$ of the differential equation $\dot{z}$ has bounded orbits. This finishes the proof.

\subsection{Proof of Theorem~\ref{thm:ftrlrec}}
\label{app_sec:ftrlrecproof}
Theorem~\ref{thm:ftrlrec} states that the {\ftrl} dynamics are Poincar\'{e} recurrent in periodic zero-sum polymatrix games. The proof of this claim follows from Lemma~\ref{lem:ftrldiv}, Lemma~\ref{lem:boundorbit} and the methods described in Section~\ref{sec:periodic}. Indeed, to begin, observe that the $\dot{z}$ dynamics given in~\eqref{eq:zdot} are $T$-periodic. This follows immediately from the definition of a $T$-periodic system as described in Section~\ref{sec:periodic} and the fact that the payoff matrices are $T$-periodic. Consider the discrete-time dynamical system defined by the Poincar\'{e} map $\phi^{T}$ that arises. This system retains the volume preservation property of the continuous-time system from Lemma~\ref{lem:ftrldiv} since as presented in Section~\ref{sec:periodic}, if a $T$-periodic system is divergence-free then the discrete-time system defined by $\phi^{T}$ is also volume preserving~\cite[3.16.B, Thm 2]{arnold2013mathematical}. Furthermore, the bounded orbits guarantee of the continuous-time system from Lemma~\ref{lem:boundorbit} imply the discrete-time system defined by $\phi^{T}$  has bounded orbits since it is a subsequence of the continuous-time system. 

Thus, we are able to apply the Poincar\'{e} recurrence theorem to the system defined by $\phi^T$ to conclude that the discrete-time system is Poincar\'{e} recurrent. This implies that the $\dot{z}$ dynamics are Poincar\'{e} recurrent since the discrete-time system defined by $\phi^T$ forms a subsequence of the continuous-time system. Finally, the Poincar\'{e} recurrence of the $\dot{z}$ dynamics directly imply the Poincar\'{e} recurrence of the {\ftrl} strategies. Indeed, since there is an increasing sequence of times $t_n$ such that $z(t_n)\rightarrow z(0)$ by Poincar\'{e} recurrence, so using continuity there is also an increasing sequence of times $t_n$ such that $x(t_n)=Q(y(t_n))=\hat{Q}(z(t_n))\rightarrow\hat{Q}(z(t_0))=x(0)$ which means the {\ftrl} dynamics are Poincar\'{e} recurrent.

\section{{\ftrl} Time-Average Convergence Results: Proofs for Section~\ref{sec:ftrltavg}}
\label{app_sec:taverage}
This appendix includes the proofs of Theorem~\ref{thm:tavgconvergence} and Proposition~\ref{prop:tarep}.

\subsection{Proof of Theorem~\ref{thm:tavgconvergence}}
\label{app_sec:ta1proof}
The outline of this proof is as follows.
We begin by restating the relevant notation specialized to periodic zero-sum bimatrix games and then provide a more formal mathematical statement of the claim being proven. Following that we introduce a technical result regarding the bounded regret property of {\ftrl} dynamics and state the implications that can be drawn from it. Finally, using the implications of bounded regret and properties of zero-sum bimatrix games we reach the conclusion.

\textbf{Notation.}
Recall that we consider a periodic zero-sum bimatrix game for this result, which is a game that consists of a pair of players $i$ and $j$ and the bimatrix game between them at time $t\geq 0$ is described by the pair of payoffs $\{A^{ij}(t), A^{ji}(t)\}=\{A(t), -A^{\top}(t)\}$ and the sequence is periodic so that $\{A^{ij}(t+T), A^{ji}(t+T)\}=\{A^{ij}(t), A^{ji}(t)\}$ or equivalently $\{A(t+T), -A^{\top}(t+T)\}=\{A(t), -A^{\top}(t)\}$ for some finite period $T$ and all time $t\geq 0$. Moreover, the bimatrix game at each time $t\geq 0$ is zero-sum which means that $u_i(x_i, x_j, t)+u_j(x_i, x_j, t)=0$ for any strategy pair $x_i\in \X_i$ and $x_j\in \X_j$. Observe that we include the time-dependence $t$ in the notation of player's utility to make it explicit the utility is time-dependent as a result of the time-varying payoff matrix. Finally, let the strategy pair $(x_i^{\ast}, x_j^{\ast})\in \X_i\times \X_j$ denote the time-invariant Nash equilibrium.

\textbf{Formal Statement of Result.} Our goal is to prove that the time-average utility of each player converges to the time-average of the values of the games over a period. That is, we seek to show 
\begin{equation}
\lim_{t\rightarrow\infty}\frac{1}{t}\int_{0}^t u_i(x(\tau),\tau)d\tau = \frac{1}{T}\int_{0}^T u_i(x_i^{\ast}, x_j^{\ast},\tau)d\tau=\bar{V}
\label{eq:statement1}
\end{equation}
and 
\begin{equation}
\lim_{t\rightarrow\infty}\frac{1}{t}\int_{0}^t u_j(x(\tau),\tau)d\tau = \frac{1}{T}\int_{0}^T u_j(x_j^{\ast}, x_i^{\ast},\tau)d\tau=-\bar{V},
\label{eq:statement2}
\end{equation}
where $\bar{V}:=\frac{1}{T}\int_{0}^TV(\tau)d\tau$ denotes the time-average of the values of the games over a period and $V(\tau)$ denotes the value of the game at time $\tau$ for any $\tau \geq 0$.

\paragraph{Bounded Regret Property.}
The proof of the above statement requires an intermediate technical result. The following result of~\citet{mertikopoulos2018cycles} states that regardless of what other players do in a polymatrix game (not necessarily zero-sum), if a player follows {\ftrl} learning dynamics then the regret of the player is bounded. It is important to remark that this result directly applies to periodic zero-sum polymatrix games. This follows from the fact that there is no assumptions on the behavior of other players, so the dynamics from the game can be viewed as arising from the behavior of the other players.
\begin{proposition}[Theorem 3.1,~\citealt{mertikopoulos2018cycles}]
Let $h_{\max, i}=\max_{x_i\in \X_i}h_i(x_i)$ and $h_{\min, i}=\min_{x_i\in \X_i}h_i(x_i)$. If player $i\in V$ in a polymatrix game follows {\ftrl} dynamics, then for every continuous trajectory of play $x_{-i}(t)$ of the opponents of player $i$ the following regret bound holds:
\begin{equation*}
\max_{x_i\in \X_i}\frac{1}{t}\int_{0}^t \big[u_i(x_i,x_{-i}(\tau), \tau)-u_i(x(\tau),\tau)\big]d\tau \leq \frac{h_{\max, i}-h_{\min, i}}{t}.
\end{equation*}
\label{prop:regret}
\end{proposition}

\textbf{Implications of Bounded Regret.}
Proposition~\ref{prop:regret} ensures that the following bounds hold for the regret of player $i$:
\begin{align}
\frac{1}{t}\int_{0}^t \big[u_i(x_i^{\ast},x_{j}(\tau),\tau)-u_i(x(\tau),\tau)\big]d\tau 
&\leq \max_{x_i\in \X_i}\frac{1}{t}\int_{0}^t \big[u_i(x_i,x_{j}(\tau),\tau)-u_i(x(\tau),\tau)\big]d\tau \notag \\
&\leq \frac{h_{\max, i}-h_{\min, i}}{t}.
\label{eq:regi}
\end{align}
Similarly, Proposition~\ref{prop:regret} guarantees the following bounds hold for the regret of player $j$:
\begin{align}
\frac{1}{t}\int_{0}^t \big[u_j(x_j^{\ast},x_{i}(\tau),\tau)-u_j(x(\tau),\tau)\big]d\tau 
&\leq \max_{x_j\in \X_j}\frac{1}{t}\int_{0}^t \big[u_j(x_j,x_{i}(\tau),\tau)-u_j(x(\tau),\tau)\big]d\tau \notag \\
&\leq \frac{h_{\max, j}-h_{\min, j}}{t}.
\label{eq:regj}
\end{align}
Observe that the lower bounds in~\eqref{eq:regi} and~\eqref{eq:regj} hold by replacing the maximizing argument over the strategy space of a player with a fixed strategy. In particular, the fixed strategy is taken to be the invariant Nash equilibrium strategy for player $i$ or $j$.

Now, taking the limit as $t\rightarrow \infty$ of each side of~\eqref{eq:regi} and using the zero-sum property of the bimatrix game at each time $\tau\geq 0$, we obtain the following:
\begin{align}
\lim_{t\rightarrow\infty}\frac{1}{t}\int_{0}^t \big[u_i(x_i^{\ast},x_{j}(\tau),\tau)-u_i(x(\tau),\tau)\big]d\tau 
&= \lim_{t\rightarrow\infty}\frac{1}{t}\int_{0}^t \big[u_i(x_i^{\ast},x_{j}(\tau),\tau)+u_j(x(\tau),\tau)\big]d\tau \notag \\
&\leq 0. \label{eq:regi1}
\end{align}
Similarly, taking the limit as $t\rightarrow \infty$ of each side of~\eqref{eq:regj} and using the zero-sum property of the bimatrix game at each time $\tau \geq 0$, we get that:
\begin{align}
\lim_{t\rightarrow\infty} \frac{1}{t}\int_{0}^t \big[u_j(x_j^{\ast},x_{i}(\tau),\tau)-u_j(x(\tau),\tau)\big]d\tau &= \lim_{t\rightarrow\infty} \frac{1}{t}\int_{0}^t \big[u_j(x_j^{\ast},x_{i}(\tau),\tau)+u_i(x(\tau),\tau)\big]d\tau \notag \\
\leq 0.
\label{eq:regj1}
\end{align}
\textbf{Time-Average Utility Convergence.} We now proceed to show that
\begin{equation}
\lim_{t\rightarrow \infty}\frac{1}{t}\int_{0}^t u_i(x_i^{\ast}, x_j^{\ast},\tau)d\tau \leq \lim_{t\rightarrow\infty}\frac{1}{t}\int_{0}^t u_i(x(\tau),\tau)d\tau \leq \lim_{t\rightarrow \infty}\frac{1}{t}\int_{0}^t u_i(x_i^{\ast}, x_j^{\ast},\tau)d\tau.
\label{eq:lowerupper}
\end{equation}
The lower bound on the time-average utility of player $i$ holds by the following analysis that is explained below:
\begin{align}
\lim_{t\rightarrow\infty}\frac{1}{t}\int_{0}^t u_i(x(\tau), \tau)d\tau &\geq \lim_{t\rightarrow\infty}\frac{1}{t}\int_{0}^t u_i(x_i^{\ast},x_{j}(\tau), \tau)d\tau \label{eq:lavgstep1}\\
&\geq \lim_{t\rightarrow\infty}\frac{1}{t}\int_{0}^t \min_{x_j\in \X_j}u_i(x_i^{\ast}, x_j, \tau)d\tau \label{eq:lavgstep2}\\
&=\lim_{t\rightarrow\infty}\frac{1}{t}\int_{0}^t u_i(x_i^{\ast}, x_j^{\ast}, \tau)d\tau. \label{eq:lavgstep3}
\end{align}
The inequality in~\eqref{eq:lavgstep1} is a direct implication of~\eqref{eq:regi1}. Moreover, the inequality in~\eqref{eq:lavgstep2} is immediate by the fact that any fixed strategy of player $j$ must give at least as much utility to player $i$ as the strategy which minimizes the utility of player $i$. Finally, the last conclusion in~\eqref{eq:lavgstep3} holds by the definition of a Nash equilibrium in a zero-sum bimatrix game.

The upper bound on the time-average utility of player $i$ holds by the following similar analysis that is detailed below:
\begin{align}
 \lim_{t\rightarrow\infty}\frac{1}{t}\int_{0}^t u_i(x(\tau), \tau)d\tau 
&\leq -\lim_{t\rightarrow\infty} \frac{1}{t}\int_{0}^t u_j(x_j^{\ast},x_{i}(\tau), \tau) d\tau \label{eq:uavgstep1} \\
&= \lim_{t\rightarrow\infty} \frac{1}{t}\int_{0}^t u_i(x_{i}(\tau),x_j^{\ast}, \tau) d\tau  \label{eq:uavgstep2} \\
&\leq \lim_{t\rightarrow\infty} \frac{1}{t}\int_{0}^t \max_{x_i\in \X_i}u_i(x_{i},x_j^{\ast}, \tau) d\tau \label{eq:uavgstep3} \\ 
&= \lim_{t\rightarrow\infty}\frac{1}{t}\int_{0}^t u_i(x_i^{\ast}, x_j^{\ast}, \tau)d\tau.
\label{eq:uavgstep4}
\end{align}
The inequality in~\eqref{eq:uavgstep1} follows directly from~\eqref{eq:regj1} and the equality in~\eqref{eq:uavgstep2} is a result of the zero-sum property of the game at each time $\tau\geq 0$. Furthermore, the inequality in~\eqref{eq:uavgstep3} holds by the fact that the the strategy of player $i$ that maximizes the utility must give at least as much utility as any fixed strategy. The last conclusion in~\eqref{eq:uavgstep4} again holds by the definition of a Nash equilibrium in a zero-sum bimatrix game. 

The preceding arguments prove that the claimed inequalities in~\eqref{eq:lowerupper} hold. Observe that the time-average of the utility values of player $i$ at the invariant Nash equilibrium converge to the time-average of the values of the games over a period as a result of the periodic nature of the game and the fact that the utility value at any Nash equilibrium in a zero-sum bimatrix game is unique. That is,
\begin{equation*}
\lim_{t\rightarrow \infty}\frac{1}{t}\int_{0}^t u_i(x_i^{\ast}, x_j^{\ast},\tau)d\tau = \frac{1}{T}\int_{0}^T u_i(x_i^{\ast}, x_j^{\ast},\tau)d\tau = \bar{V}.
\end{equation*}
Thus, the squeeze theorem applied to~\eqref{eq:lowerupper} allows us to conclude the statement given in~\eqref{eq:statement1}:
\begin{equation*}
\lim_{t\rightarrow\infty}\frac{1}{t}\int_{0}^t u_i(x(\tau),\tau)d\tau = \frac{1}{T}\int_{0}^T u_i(x_i^{\ast}, x_j^{\ast},\tau)d\tau = \bar{V}
\end{equation*}
Finally, by the zero-sum property of the bimatrix game at each time $\tau\geq 0$, the statement given in~\eqref{eq:statement2} immediately follows from the equation above. That is,
\begin{equation*}
\lim_{t\rightarrow\infty}\frac{1}{t}\int_{0}^t u_j(x(\tau),\tau)d\tau = \frac{1}{T}\int_{0}^T u_i(x_j^{\ast}, x_i^{\ast},\tau)d\tau = -\bar{V}.
\end{equation*}
This finishes the proof.

\subsection{Proof of Proposition~\ref{prop:tarep}}
\label{app_sec:ta2proof}
We now provide the proof of Proposition~\ref{prop:tarep} stating that there exists periodic zero-sum bimatrix games satisfying Definition~\ref{def:tvpolymatrix} in which the time-average strategies of {\ftrl} dynamics fail to converge to the time-invariant Nash equilibrium. 

To prove this result we construct a specific periodic zero-sum bimatrix game that is the basis of the counterexample. In order to demonstrate that the {\ftrl} strategies may not converge to the time-invariant Nash equilibrium, we consider the regularization function that leads to the replicator dynamics. Then, for replicator dynamics in the constructed game, we prove that the strategies are symmetric about the half period of the game and consequently return to the initial condition in a period of the game so that the time-average of the strategies in the limit corresponds to the time-average of the strategies over a half-period of the game. Finally, we use this property to show that the choice of the period of the game can ensure that the time-average strategies cannot converge to the time-invariant Nash equilibrium.

\textbf{Counterexample Construction.}
A periodic zero-sum bimatrix game between players $i$ and $j$ is described by a periodic sequence of payoffs where the game at time $t\geq 0$ is described by the pair of payoffs $\{A^{ij}(t), A^{ji}(t)\}=\{A(t), -A^{\top}(t)\}$.
To obtain a counterexample, we consider the periodic zero-sum bimatrix game defined by
\begin{equation*}
A(t)=\gamma(t)A\quad \text{where} \quad A=\begin{bmatrix}\ \ \ 1&-1\\-1&\ \ 
 \ 1 \end{bmatrix} \quad \text{and} \quad \gamma(t)=\sin\Big(\frac{2\pi t}{T}\Big).
\end{equation*}
This game corresponds to a periodic version of matching pennies and the period of the game is $T$. The joint strategy $(x_i^{\ast}, x_j^{\ast})$ where $x_i^{\ast}=(1/2, 1/2)$ and $x_j^{\ast}=(1/2, 1/2)$ is the unique time-invariant Nash equilibrium of the game.

Recall that in periodic zero-sum bimatrix games between players $i$ and $j$, we denote the utility of each player at time $t\geq 0$ under the joint strategy $x(t)$ by $u_i(x(t), t)$ and $u_j(x(t), t)$ to emphasize the dependence on the time-dependent payoffs which are given by $\{\gamma(t)A, -\gamma(t)A^{\top}\}$ in this construction. Furthermore, in this problem construction we denote the utility of each player with payoffs $\{A, -A^{\top}\}$ under the joint strategy $x(t)$ by $u_i(x(t))$ and $u_j(x(t))$ where $A$ is defined at the matching pennies payoff matrix defined above.

The regularization function $h_i(x_i)=\sum_{\alpha\in \A_i}x_{i\alpha}\log x_{i\alpha}$ in {\ftrl} dynamics gives rise to the replicator dynamics commonly studied in evolutionary game theory. 
For the periodic zero-sum bimatrix game under consideration, the replicator dynamics for any strategy $\alpha\in \A_i$ of player $i$ are given by
\begin{equation*}
\dot{x}_{i\alpha}(t) = x_{i\alpha}(t)\big[u_{i\alpha}(x(t), t)-u_i(x(t),t)\big] = \gamma(t)x_{i\alpha}(t)\big[u_{i\alpha}(x(t))-u_i(x(t))\big]:=\gamma(t) \dot{x}_{i\alpha}'(t).
\end{equation*}
Similarly, the replicator dynamics for any strategy $\alpha\in \A_j$ of player $j$ are given by
\begin{equation*}
\dot{x}_{j\alpha}(t) = x_{j\alpha}(t)\big[u_{j\alpha}(x(t), t)-u_j(x(t),t)\big] = \gamma(t)x_{j\alpha}(t)\big[u_{j\alpha}(x(t))-u_j(x(t))\big]:=\gamma(t) \dot{x}_{j\alpha}'(t).
\end{equation*}
We now analyze the time-average of the dynamics of any strategy for each player in this game and show that they do not correspond to the time-invariant Nash equilibrium.

\textbf{Time-Average Strategies.}
We begin by showing that for each player $k\in \{i, j\}$ and strategy of the player $\alpha\in \A_k$,
\begin{equation}
x_{k\alpha}\big(\tfrac{T}{2}+t\big)=x_{k\alpha}\big(\tfrac{T}{2}-t\big).
\label{eq:halfperiod}
\end{equation}
To see this, observe that for each player $k\in \{i, j\}$ and strategy of the player $\alpha\in \A_k$ and some initial condition $t_0$,
\begin{equation}
x_{k\alpha}(t)=x_{k\alpha}(t_0)+\int_{t_0}^t\dot{x}_{j\alpha}(\tau) d\tau=x_{k\alpha}(t)+\int_{t_0}^t\sin\big(\tfrac{2\pi}{T}\tau\big)\dot{x}_{k\alpha}'(\tau) d\tau.
\label{eq:halfperiod2}
\end{equation}
To prove the claim in~\eqref{eq:halfperiod}, we show that both $x_{k\alpha}\big(\tfrac{T}{2}+t\big)$ and $x_{k\alpha}\big(\tfrac{T}{2}-t\big)$ satisfy the same ordinary differential equation and initial condition. That is, we  invoke the fundamental theorem of ordinary differential equations which says the solutions exist and are unique so that the claim holds. 

Indeed, from~\eqref{eq:halfperiod2} and the fundamental theorem of calculus
\begin{align*}
\frac{d}{dt}x_{k\alpha}\big(\tfrac{T}{2}+t\big)
&=\dot{x}_{k\alpha}'\big(\tfrac{T}{2}+t\big)\sin\big(\tfrac{2\pi}{T}\big(\tfrac{T}{2}+t\big)\big)\\
&=\dot{x}_{k\alpha}'\big(\tfrac{T}{2}+t\big)\sin\big(\pi+\tfrac{2\pi}{T}t\big)\\
&=-\dot{x}_{k\alpha}'\big(\tfrac{T}{2}+t\big)\sin\big(\tfrac{2\pi}{T}t\big).
\end{align*}
Similarly,
\begin{align*}
\frac{d}{dt}x_{k\alpha}\big(\tfrac{T}{2}-t\big)
&=-\dot{x}_{k\alpha}'\big(\tfrac{T}{2}-t\big)\sin\big(\tfrac{2\pi}{T}\big(\tfrac{T}{2}-t\big)\big)\\
&=-\dot{x}_{k\alpha}'\big(\tfrac{T}{2}-t\big)\sin\big(\pi-\tfrac{2\pi}{T}t\big)\\
&=-\dot{x}_{k\alpha}'\big(\tfrac{T}{2}-t\big)\sin\big(\tfrac{2\pi}{T}t\big).
\end{align*}
We conclude that the functions $x_{k\alpha}\big(\tfrac{T}{2}+t\big)$ and $x_{k\alpha}\big(\tfrac{T}{2}-t\big)$  satisfy the same ordinary differential equation. That is, the functional form of the ordinary differential equation is the same in both expressions.
Furthermore, $x_{k\alpha}\big(\tfrac{T}{2}+0\big)=x_{k\alpha}\big(\tfrac{T}{2}-0\big)=x_{k\alpha}\big(\tfrac{T}{2}\big)$ so they satisfy the same initial condition. Hence, invoking the uniqueness property of the fundamental theorem of ordinary differential equations, the claim given in~\eqref{eq:halfperiod} holds.

The property in~\eqref{eq:halfperiod} implies for each player $k\in \{i, j\}$ and strategy of the player $\alpha\in \A_k$ that
\begin{equation*}
\lim_{t\rightarrow \infty}\frac{1}{t}\int_0^t x_{k\alpha}(\tau) d\tau=\frac{1}{T}\int_0^T x_{k\alpha}(\tau) d\tau=\frac{2}{T}\int_0^{T/2} x_{k\alpha}(\tau) d\tau.
\end{equation*}
That is, the limiting time-average strategy is equal to the time-average strategy over half a period of the periodic game.

Now recall that the time-invariant Nash equilibrium strategy is given by the joint strategy $(x_i^{\ast}, x_j^{\ast})$ where $x_i^{\ast}=(1/2, 1/2)$ and $x_j^{\ast}=(1/2, 1/2)$. Thus, to finish the proof we need to show for some player $k\in \{i, j\}$ and strategy of $\alpha\in \A_k$ that
\begin{equation*}
\frac{2}{T}\int_0^{T/2} x_{k\alpha}(\tau) d\tau\neq \frac{1}{2}.
\end{equation*}
To see that the claim above holds, recall from~\eqref{eq:halfperiod2} that 
\begin{equation*}
x_{k\alpha}(t)=x_{k\alpha}(0)+\int_{0}^t\sin\big(\tfrac{2\pi}{T}\tau\big)\dot{x}_{k\alpha}'(\tau) d\tau.
\end{equation*}
Observe that for any $\tau\geq 0$,
\[-1\leq \sin\Big(\frac{2\pi}{T}\tau\Big)\leq 1 \quad \text{and}\quad -2\leq \dot{x}_{k\alpha}'(\tau)\leq 2.\]
The previous expressions combine to imply
\begin{equation*}
x_{k\alpha}(0)-2t\leq x_{k\alpha}(t)\leq x_{k\alpha}(0)+2t
\end{equation*}
and consequently
\begin{equation*}
x_{k\alpha}(0)- \frac{T}{2}\leq \frac{2}{T}\int_0^{T/2} x_{k\alpha}(\tau) d\tau \leq x_{k\alpha}(0)+\frac{T}{2}.
\end{equation*}
Finally, suppose that $x_{k\alpha}(0)>1/2$. Then, when $T< 2(x_{k\alpha}(0)-1/2)$, the time-average of the strategy \[\frac{2}{T}\int_0^{T/2} x_{k\alpha}(\tau) d\tau>\frac{1}{2}.\] Analogously, suppose that $x_{k\alpha}(0)<1/2$. Then, when $T<2(1/2-x_{k\alpha}(0))$, the time-average of the strategy \[\frac{2}{T}\int_0^{T/2} x_{k\alpha}(\tau) d\tau<\frac{1}{2}.\] 
Thus, unless the players initialize at the time-invariant Nash equilibrium strategy, there is a choice of the period $T$ of the periodic zero-sum matrix such that the time-average of the strategies do not converge to the time-invariant Nash equilibrium.
This completes the proof.

\section{Experiments}\label{app_sec:experiments}
In this section, we present additional simulations and details that serve to strengthen the results in the main paper. 

\subsection{{\gda} Results}
First, for continuous-time {\gda} dynamics we show that Poincar\'e recurrence holds in a periodic zero-sum bilinear game. We consider the ubiquitous Matching Pennies game with payoff matrix $A=\bigl(\begin{smallmatrix}1 & -1\\ -1 & 1\end{smallmatrix} \bigr)$. We then use the following periodic rescaling with period $2\pi$:
 \begin{align}\label{eqn:rescaling} \alpha(t) = \left\{
\begin{array}{ll}
      \sin(t) & 0\leq t \leq \frac{3\pi}{2} \\
      \left(\frac{2}{\pi}\right)(t\ \mathrm{mod}(2\pi)-2\pi) & \frac{3\pi}{2} \leq t\leq 2\pi
\end{array} 
\right. 
\end{align}
Hence, the bilinear zero-sum game at time $t\geq 0$ is then described by the payoffs $\{\alpha(t)A(t), -\alpha(t)A(t)^{\top}\}$.
When agents use {\gda} learning dynamics, we see from Figure \ref{fig:gda_1} that the agents' trajectories when plotted alongside the value of the periodic rescaling are bounded. 

\begin{figure}[!htb]
    \centering
    \subfigure{\includegraphics[scale=0.3]{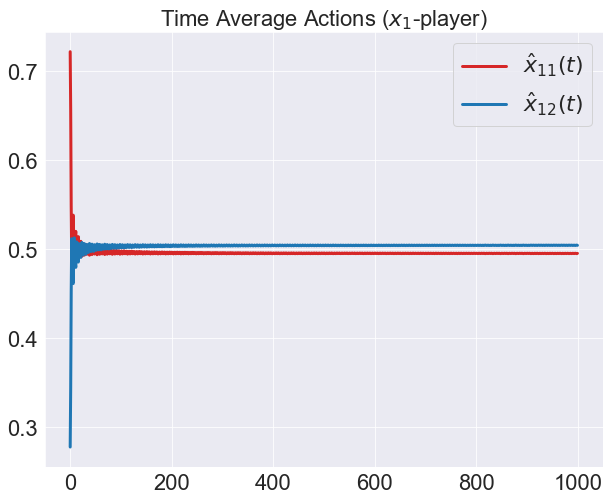}}\quad
    \subfigure{\includegraphics[scale=0.305]{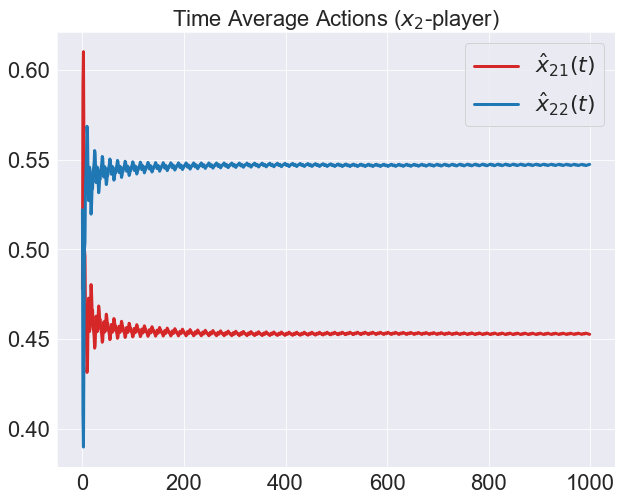}}\quad
    \caption{Time-average convergence away from the mixed Nash}
    \label{fig:counterexample}
\end{figure}

Another key result in the space of {\gda} learning dynamics is that the time-average behavior fails to converge in general to the time invariant equilibrium $(0,0)$. A simple counterexample can be constructed by considering a periodic zero-sum bilinear game with $x_1,x_2\in \mb{R}$ and a periodic rescaling $\beta(t)$ such that $\beta(t)=\beta(t+T)$ with $T=3\pi$ for any $t\geq 0$. 
Moreover, let the rescaling evolve over a period as follows:
 \begin{align} \beta(t) = \left\{
\begin{array}{ll}
      -1 & 0\leq t \leq \pi \\
      \ \ \ 1 & \pi \leq t\leq \frac{3\pi}{2}\\
      -1 &  \frac{3\pi}{2}\leq t\leq 3\pi.
\end{array} 
\right. \end{align}
For the simulation, we consider the payoff matrices described by $\{\beta(t)A, -\beta(t)A^{\top}\}$ where $A=\bigl(\begin{smallmatrix}1 & -1\\ -1 & 1\end{smallmatrix} \bigr)$. We show for this example that when both players use {\gda}, the time average strategy of each player remains bounded away from the Nash $[1/2, 1/2]$, as shown in Figure \ref{fig:counterexample}.

\subsection{{\ftrl} Results}

\begin{figure}[!htb]
    \centering
    \includegraphics[scale=0.35]{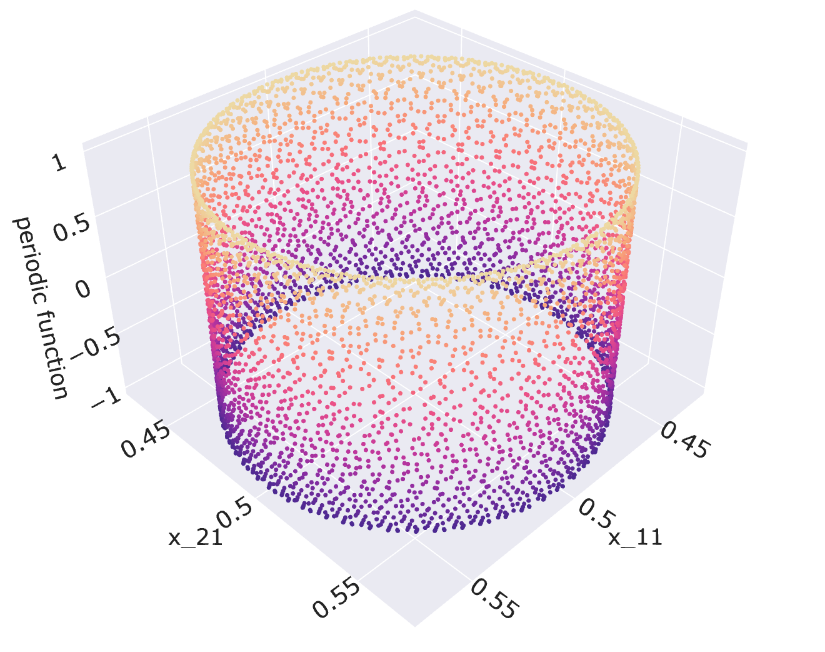}
    \caption{Periodically Rescaled Matching Pennies (Replicator)}
    \label{fig:rd_1}
\end{figure}

For the case of {\ftrl} dynamics, we perform simulations on Matching Pennies updated with replicator dynamics. The reader is reminded of the definition of replicator dynamics as a continuous analogue of multiplicative weights update, as described in Section \ref{subsec:replicator}. In polymatrix games, replicator dynamics for each $i\in V$ uses regularization function $h_i(x_i)=\sum_{\alpha\in \A_i}x_{i\alpha}\log x_{i\alpha}$ in the {\ftrl} dynamics. Like the {\gda} case, we also use the periodic rescaling described in Equation \ref{eqn:rescaling} and obtain recurrent dynamics, as seen in Figure \ref{fig:rd_1}. 

\begin{figure}[!htb]
    \centering
    \subfigure{\includegraphics[scale=0.3]{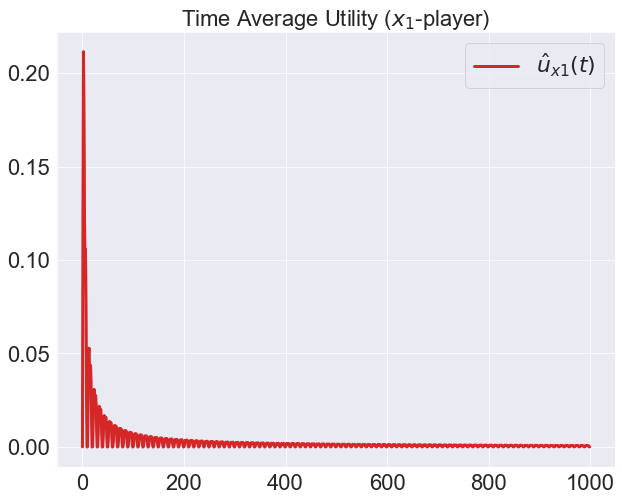}}\quad
    \subfigure{\includegraphics[scale=0.3]{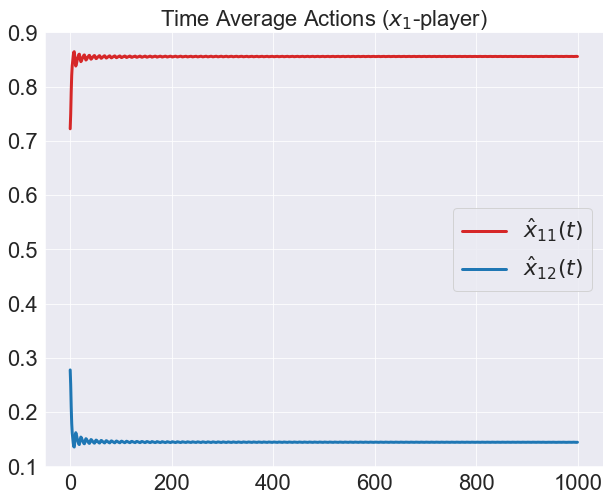}}\quad
    \caption{Time average results for MP rescaled with $sin$ function. Notice that although the average utility goes to 0, the time average actions/strategy does not go to the time invariant Nash [1/2, 1/2].}
    \label{fig:tarep}
\end{figure}

Theorem \ref{thm:tavgconvergence} states that the time-average utility of each player converges to the time-average value of periodic zero-sum games when each player follows {\ftrl} dynamics. However, Proposition \ref{prop:tarep} states that there exist periodic zero-sum bimatrix games where the time-average strategies of {\ftrl} dynamics fail to converge to the time-invariant Nash equilibrium. Here, we show a simple example that exhibits both results. Consider a Matching Pennies game that is rescaled with a $\sin$ function. Specifically, the periodic bimatrix game is given by 
\begin{align} \label{eqn:singame}
    A(t) = \sin(t)\begin{bmatrix}\ \ \ 1&-1\\-1&\ \ 
 \ 1 \end{bmatrix} 
\end{align}
Even with this simple example, the time-average utilities for both players go to zero, while the time-average strategies do not converge to the $[1/2, 1/2]$ time-invariant Nash, as seen in Figure \ref{fig:tarep}.

\subsection{Large Scale Simulation}
\begin{figure}[!ht]
  \centering
  {\includegraphics[scale=0.42]{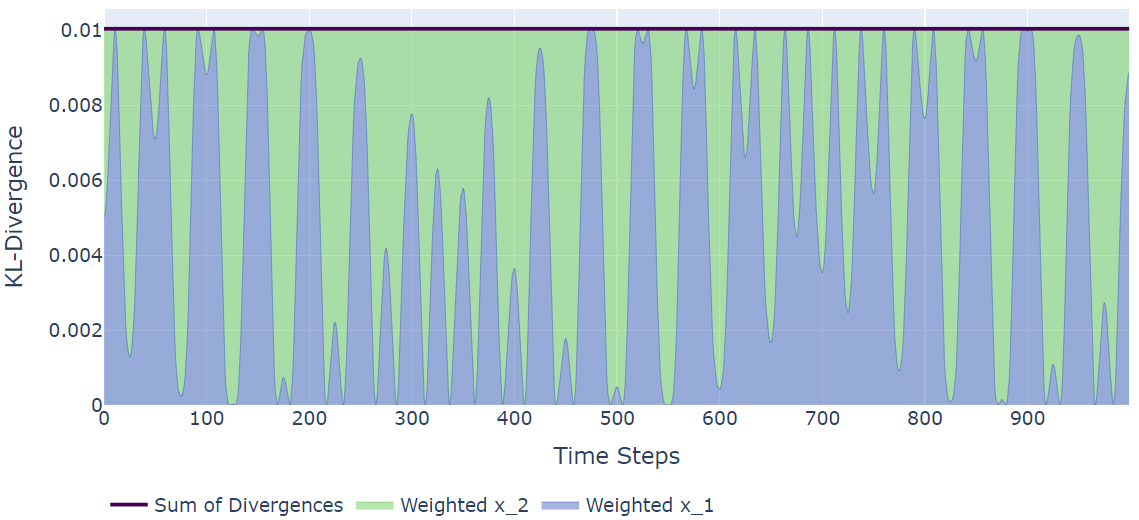}}
  \caption{Time invariant function for two player periodically rescaled Matching Pennies game \label{fig:kldiv1}}
\end{figure}

We now perform simulations on larger-scale systems that present our findings in a more visually striking manner. Firstly, the time-invariant function presented in Lemma \ref{lem:ftrldiv} and its proof can be demonstrated in any two-player periodic zero-sum polymatrix game. For replicator dynamics, the invariant function is the KL-divergence between each player's strategy and the unique mixed Nash. Using the same simulated data that was used to generate Figure \ref{fig:rd_1}, we show that the sum of divergences is indeed constant when both agents are using replicator dynamics. Figure \ref{fig:kldiv1} shows this phenomenon in the case where there are just two players playing a periodically rescaled Matching Pennies game. Specifically, the blue area represents the KL-divergence of the first player from the mixed Nash over time, and the green area represents the divergence of the second player.

\begin{figure}[!htb]
    \centering
    \subfigure{\includegraphics[width=0.9\textwidth]{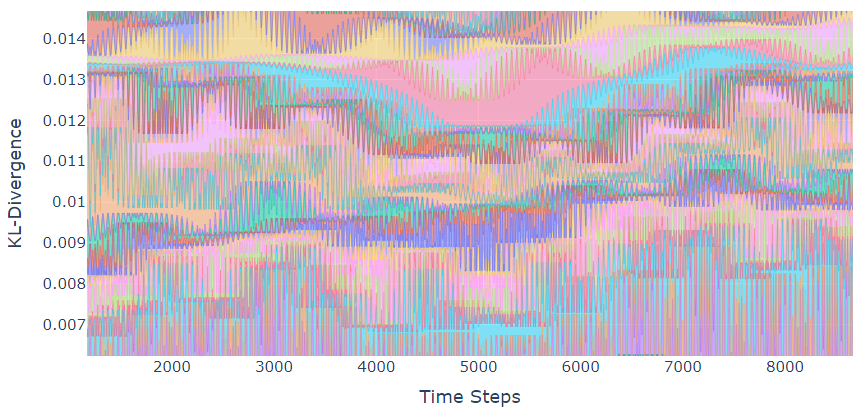}}\quad
    \subfigure{\includegraphics[width=0.91\textwidth]{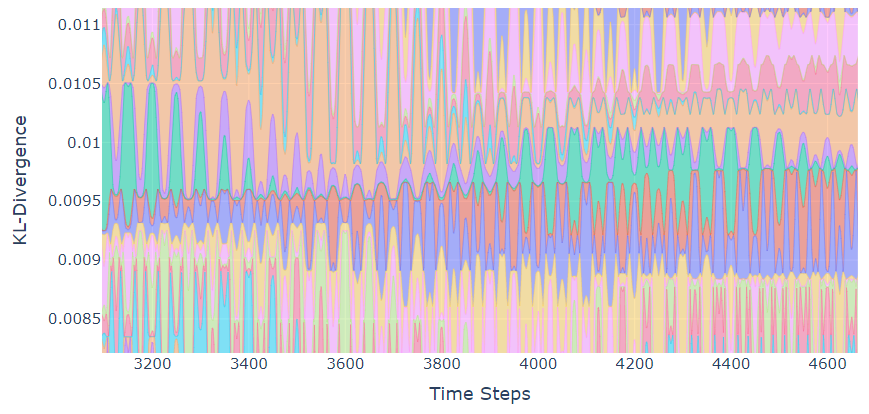}}\quad
    \caption{Zoomed-in time invariant functions for 64-player game.}
    \label{fig:kldiv_zoom}
\end{figure}

To extend this formulation to the multiplayer setting, we implement a graphical polymatrix game where a number of agents are arranged in a line. Each agent then plays a bimatrix game against the agent directly adjacent to them, and the final agent also plays against the first agent. This results in a `toroid'-like chain of games, where each agent plays against two other agents. In our simulation, each pair of agents plays the Matching Pennies game rescaled with $\sin$ against each other (Equation \ref{eqn:singame}). With this system, we simulated a chain of games with 64 nodes (agents), resulting in much more complex dynamics than the two player case. Indeed, in Figure~\ref{fig:kldiv_zoom} we show zoomed-in plots of Figure~\ref{fig:kldiv}. Here, similar to previous plots, each player's KL-divergence is represented by a different-colored area. The figure showcases that on a more granular level, the individual KL-divergences of each agent can become extremely erratic, and look nowhere near periodic. Nevertheless, we see from Figure \ref{fig:kldiv} that the sum of KL-divergences remains constant.

\begin{figure}[!ht]
  \centering
  {\includegraphics[scale=0.4]{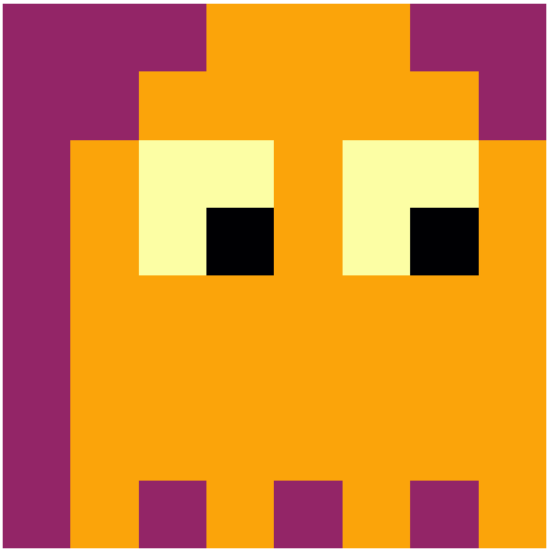}}
  \caption{$8 \times 8$ grid of colors generated by sigmoid function \label{fig:clyde}}
\end{figure}

We also represent the trajectories of each agent by equating the strategy values of each player to RGB values in an 8x8 grid. In particular, the color of each pixel on the grid represents the probability of the respective player playing the first strategy, tuned with a sigmoid function. We then select initial conditions the correspond to RGB values such that they form the image shown in Figure \ref{fig:clyde}. With the sigmoid function, any changes from the initial condition are reflected by changes in the color of the individual pixels. Thus, as agents play pairwise bimatrix games using replicator dynamics, the colors of the grid evolve. If recurrence holds, we expect to see the same image after some time. For the case of the Matching Pennies games rescaled with $\sin$, we obtain the various images found in Figure \ref{fig:clydemain}, which exhibit recurrence. We also provide an animation showing how the figure evolves over time in the supplementary Jupyter notebook.

\subsection{Reproducibility Details}
All experiments performed for this work were done using Python 3.7 and have been compiled into a Jupyter notebook for ease of viewing. Running the code requires only basic scientific computing packages such as NumPy and SciPy, as well as data visualization packages such as Matplotlib and Plotly. Most of the code in our submission has been edited such that it can be easily executed on a standard computer in a matter of minutes.

\newpage
\section*{Appendix References}
\medskip

{

[1] Alexander, J.A.\ \& Mozer, M.C.\ (1995) Template-based algorithms for
connectionist rule extraction. In G.\ Tesauro, D.S.\ Touretzky and T.K.\ Leen
(eds.), {\it Advances in Neural Information Processing Systems 7},
pp.\ 609--616. Cambridge, MA: MIT Press.

[2] Bower, J.M.\ \& Beeman, D.\ (1995) {\it The Book of GENESIS: Exploring
  Realistic Neural Models with the GEneral NEural SImulation System.}  New York:
TELOS/Springer--Verlag.

[3] Hasselmo, M.E., Schnell, E.\ \& Barkai, E.\ (1995) Dynamics of learning and
recall at excitatory recurrent synapses and cholinergic modulation in rat
hippocampal region CA3. {\it Journal of Neuroscience} {\bf 15}(7):5249-5262.
}

\end{document}